%% file: EXO-18-002_temp.tex
\pdfoutput=1
\documentclass[11pt,twoside,a4paper,cmspaper,final,collab]{cms-tdr}
\def\svnVersion{10a2ebc}\def\svnDate{2025/04/24}\def\cmsCernNoTag{CERN-EP-2024-254}\def\cmsCernDate{\today}\def\cmsMessage{Published in the Journal of High Energy Physics as \href{http://dx.doi.org/10.1007/JHEP04(2025)109}{\doi{10.1007/JHEP04(2025)109}.}}
\begin{document}\cmsNoteHeader{EXO-18-002}

\newlength\cmsFigWidth
\setlength\cmsFigWidth{0.4\textwidth}

\newcommand{\rhadrons}{R-hadrons}
\newcommand{\rhadron}{R-hadron}
\newcommand{\stau}{\PSGt}
\newcommand{\stauL}{\PSGtL}
\newcommand{\stauR}{\PSGtR}
\DeclareRobustCommand{\PSGtLR}{{\HepSusyParticle{\PGt}{L/R}{}}\Xspace}
\newcommand{\stauLR}{\PSGtLR}
\newcommand{\stopq}{\PSQt}
\newcommand{\squark}{\PSQ}
\newcommand{\gstrip}{\ensuremath{G_{\text{i}}^{\text{Strips}}}}
\newcommand{\iasp}{\ensuremath{I_{as}^\prime}\xspace}
\newcommand{\fpix}{\ensuremath{F_{\text{i}}^{\text{Pixels}}}}
\newcommand{\ih}{\ensuremath{I_{\text{h}}}}
\newcommand{\Ih}{\ensuremath{I_{\text{h}}}}
\newcommand{\Irel}{\ensuremath{I_{\text{PF}}^{\text{rel}}}}
\newcommand{\pthscp}{\ensuremath{\pt^{\text{HSCP}}}}
\newcommand{\DeltaRmini}{\ensuremath{\Delta R_{\text{mini-iso}}}}
\newcommand{\Itrk}{\ensuremath{I_{\text{trk}}}}
\newcommand{\dedx}{\ddinline{E}{x}\xspace}
\DeclareRobustCommand{\ZPrimepsi}{{\HepParticle{\PZ}{\PGy}{\prime}}\Xspace}
\DeclareRobustCommand{\ZPrimeSSM}{{\HepParticle{\PZ}{\mathrm{SSM}}{\prime}}\Xspace}
\newcommand{\Cgrav}{\ensuremath{C_\text{grav}}\xspace}
\newcommand{\sigptoverptsq}{\ensuremath{\sigma_{\pt}/\pt^2}\xspace}
\newcommand{\sigmaptoverpt}{\ensuremath{\sigma_{\pt}/\pt}\xspace}
\newcommand{\RPF}{\ensuremath{R_{\text{P/F}}}\xspace}
\newcommand{\RPFj}{\ensuremath{R_{\text{P/F}}(j)}\xspace}
\newcommand{\Iconetrk}{\ensuremath{I_{\text{trk}}^{\Delta R< 0.3}}\xspace}
\newcommand{\dz}{\ensuremath{d_{\mathrm{z}}}\xspace}
\newcommand{\dxy}{\ensuremath{d_{\mathrm{xy}}}\xspace}
\newcommand{\NFAIL}{\ensuremath{N_{\text{FAIL}}^{\text{bkg}}(j)}}
\newcommand{\NPASS}{\ensuremath{N_{\text{PASS}}^{\text{bkg}}(j)}}

\newlength\cmsTabSkip\setlength{\cmsTabSkip}{1ex}

\cmsNoteHeader{EXO-18-002} 
\title{Search for heavy long-lived charged particles with large ionization energy loss 
in proton-proton collisions at \texorpdfstring{$\sqrt{s} = 13\TeV$}{sqrt(s) = 13 TeV}}

\date{\today}

\abstract{
A search for heavy, long-lived, charged particles with large ionization energy loss within the silicon tracker of the CMS experiment is presented. A data set of proton-proton collisions at a center of mass energy at $\sqrt{s} = 13\TeV$, collected in 2017 and 2018 at the CERN LHC, corresponding to an integrated luminosity of 101\fbinv, is used in this analysis. Two different approaches for the search are taken. A new method exploits the independence of the silicon pixel and strips measurements, while the second method improves on previous techniques using ionization to determine a mass selection. No significant excess of events above the background expectation is observed. The results are interpreted in the context of the pair production of supersymmetric particles, namely gluinos, top squarks, and tau sleptons, and of the Drell--Yan pair production of fourth generation (\PGtpr) leptons with an electric charge equal to or twice the absolute value of the electron charge ($e$). An interpretation of a \PZpr boson decaying to two \PGtpr leptons with an electric charge equal to $2e$ is presented for the first time. The 95\% confidence upper limits on the production cross section are extracted for each of these hypothetical particles.
}

\hypersetup{%
    pdfauthor={CMS Collaboration},
    pdftitle={Search for heavy long-lived charged particles with large ionization energy loss in proton-proton collisions at sqrt(s) = 13 TeV},
    pdfsubject={CMS},
    pdfkeywords={CMS, long lived, exotica, energy loss in silicon}
}

\maketitle 

\section{Introduction}

Many extensions of the standard model (SM) include heavy, long-lived, charged particles that could be produced at the CERN LHC with speeds significantly less than the speed of light~\cite{Drees:1990yw, Fairbairn:2006gg, Bauer:2009cc}, and/or electric charge $Q$ not equal to  $\pm 1e$~\cite{Kusenko:1997si, Koch:2007um,Schwinger:1966nj, Fargion:2005ep,Alimena:2019zri, Lee:2018pag}, with $e$ being the absolute value of the electron charge.
If particles predicted by these extensions beyond the SM (BSM) are produced at the LHC and they have lifetimes greater than 
$\mathcal{O}(\unit{ns})$, they could travel far in the CMS detector before decaying.
Large lifetimes are the consequence of small widths and, indeed, BSM particles can have small total decay widths as the result of small couplings to their decay products, suppressed phase space, or large mediator mass scales.
If such a particle carries electric charge, it can be generically referred to as a heavy stable charged particle (HSCP), even 
though it may actually be metastable.

For HSCPs with mass greater than about 100\GeV, a significant fraction of BSM particles produced at the LHC will have a relative velocity, $\beta \equiv v/c$, significantly smaller than 1.  
Such particles can be separated from ultra-relativistic lower-mass SM particles because of their higher rate of ionization energy loss (\dedx).
This paper describes a search for HSCPs using measurements of \dedx in the CMS pixel and silicon strip detectors.

The dependence of \dedx on particle momentum is described by the Bethe--Bloch formula~\cite{pdg}. 
Searching for candidates at large momentum with larger \dedx deposits gives sensitivity to slow, massive particles with $\abs{Q} = 1e$ and particles with $\abs{Q} > 1e$,
as the SM particles have nearly uniform collected charge per unit of path length at the 
minimum ionizing particle (MIP) level (with a mean of $\approx$3.9\unit{MeV/cm} in silicon).

Previous collider searches for HSCPs were performed at the CERN LEP collider~\cite{Barate:1997dr, Abreu:2000tn,Achard:2001qw,Abbiendi:2003yd}, DESY HERA collider~\cite{Aktas:2004pq}, the Fermilab Tevatron~\cite{Aaltonen:2009kea, Abazov:2008qu,Abazov:2011pf,Abazov:2012ina},
and the LHC~\cite{Aad:2011hz, Aad:2011yf, Aad:2011mb,Aad:2012pra, Aad:2013pqd, ATLAS:2014fka, Aad:2015oga, Aad:2015qfa, Aaboud:2016dgf, Aaboud:2016uth,Khachatryan:2011ts, Chatrchyan::2012dr, Chatrchyan:2012sp, Khachatryan:2016sfv, ATLAS_excess}, each placed upper cross section limits on various HSCP models. 
The most recent previous CMS results based on data recorded in 2015~\cite{Khachatryan:2016sfv} place lower limits at 95\% confidence level (\CL) on the mass of gluinos, top squarks, and pair-produced tau sleptons.
An analysis~\cite{ATLAS_excess} based on tracker \dedx information only, performed by the ATLAS Collaboration,
reports an excess of seven events, where $0.7 \pm 0.4$ are expected, with a local (global) significance of 3.6 (3.3) standard deviations. 
The majority of these events are reconstructed as muons candidates. However, the $\beta$ values measured in the muon chambers or in the calorimeter for these seven events
are compatible with 1, indicating that none of them are moving significantly slower than the speed of light.

The analysis presented here is a model-independent search for singly and doubly charged HSCPs 
using data collected with the CMS detector at $\sqrt{s} = 13\TeV$ in 2017--2018  corresponding to an integrated luminosity of  101\fbinv. 
In addition to an event sample significantly larger than the one in Ref.~\cite{Khachatryan:2016sfv}, this search 
benefits from the development of a new ionization discriminator that uses the charge collected in the new Phase-1 pixel detector~\cite{Phase1Pixel}, 
as well as two data-driven methods to estimate the background contribution. 

The paper is organized as follows. The CMS detector and event reconstruction are presented in Section~\ref{sec:detDesc}.  
Section~\ref{sec:signals} describes the different models considered and the associated signal samples,
while Section~\ref{sec:variables} introduces the ionization variables used in this search. 
The event selection is detailed in Section~\ref{sec:presel}, followed by the description of the two different background estimation methods in Section~\ref{sec:bkgpred}.
Section~\ref{sec:systematics} presents the systematic uncertainties associated with the signals 
and results are shown in Section~\ref{sec:results}. The paper is summarized in Section~\ref{sec:summary}.

Tabulated results are provided in the HEPData record for this analysis~\cite{hepdata}.

\section{The CMS detector and event reconstruction}
\label{sec:detDesc}

The central feature of the CMS apparatus is a superconducting solenoid of 6\unit{m} internal diameter, providing a magnetic field of 3.8\unit{T}. 
Within the solenoid volume are a silicon pixel and silicon strip tracker, a lead tungstate crystal electromagnetic calorimeter (ECAL), and a brass and scintillator hadron calorimeter (HCAL), each composed of a barrel and two endcap sections. 
Forward calorimeters extend the pseudorapidity ($\eta$) coverage provided by the barrel and endcap detectors. 
Muons are measured in gas-ionization detectors embedded in the steel flux-return yoke outside the solenoid. 
A more detailed description of the CMS detector, together with a definition of the coordinate system used and the relevant kinematic variables, can be found in Ref.~\cite{CMS:2008xjf}.

At the start of 2017, a new silicon pixel detector was installed~\cite{Phase1Pixel}; the upgraded tracker measures particles up to $\abs{\eta} < 3.0$.
During the LHC running period when the data used in this paper were recorded, the silicon tracker consisted of 1856 silicon pixel and 15\,148 silicon strip detector modules. 
In this paper, to ensure a uniform detector performance, we focus on \dedx measurements in the barrel part of the tracker, composed of four layers in the silicon pixel detector 
and 10 layers in the silicon strip detector (among which four of them are stereo, providing two separate \dedx measurements per layer).
Isolated particles of \pt = 100\GeV emitted at $\abs{\eta} < 1.4$ have track resolutions of 2.8\% in \pt and 10 (30)\mum in the transverse (longitudinal) impact parameter~\cite{Chatrchyan:1704291}.
The primary vertex is taken to be the vertex corresponding to the hardest scattering in the event, evaluated using tracking information alone, as described in Section 9.4.1 of Ref.~\cite{CMS-TDR-15-02}.

Events of interest are selected using a two-tiered trigger system. 
The first level, composed of custom hardware processors, uses information from the calorimeters and muon detectors to select events at a rate of around 100\unit{kHz} within a fixed latency of about 4\mus~\cite{CMS:2020cmk}. 
The second level, known as the high-level trigger, consists of a farm of processors running a version of the full event reconstruction software optimized for fast processing, and reduces the event rate to around 1\unit{kHz} before data storage~\cite{CMS:2016ngn}.

A particle flow (PF) algorithm~\cite{CMS:2017yfk} aims to reconstruct and identify each individual particle in an event, with an optimized combination of information from the various elements of the CMS detector. 
The energy of photons is obtained from the ECAL measurement. 
The energy of electrons is determined from a combination of the electron momentum at the primary interaction vertex as determined by the tracker, the energy of the corresponding ECAL cluster, and the energy sum of all bremsstrahlung photons spatially compatible with originating from the electron track.
The energy of muons is obtained from the curvature of the corresponding track. 
The energy of charged hadrons is determined from a combination of their momentum measured in the tracker and the matching ECAL and HCAL energy deposits, corrected for the response function of the calorimeters to hadronic showers. 
Finally, the energy of neutral hadrons is obtained from the corresponding corrected ECAL and HCAL energies.

\section{Simulated samples}
\label{sec:signals}

This model-independent search is sensitive to a wide variety of new physics models that produce long-lived, charged particles. Several signal models are used to interpret the results.
The \PYTHIA generator v8.240~\cite{Sjostrand:2014zea}, with the tune CP5~\cite{tune} and the parton distribution function set NNPDF3.1~\cite{nnpdf},  is used to produce the Monte Carlo (MC) signal samples at $\sqrt{s}=13\TeV$. 

The first type of signal consists of HSCPs that are charged under the strong force and live longer than typical hadronization scales. Such particles will therefore hadronize into a color-singlet hadronic state by picking up SM quarks or gluons. 
The pair production of long-lived gluinos (\PSg) is a common signature of mini-split supersymmetric (SUSY) models~\cite{Arvanitaki_2013} where the decay of the \PSg
into a scalar squark \squark, a supersymmetric partner of an SM quark, is forbidden because the \squark mass is much higher than the \PSg mass. The \PSg decay can then only occur via a highly virtual \squark, leading to a long-lived \PSg~\cite{Lee:2018pag}. In this analysis, a quasi-stable \PSg is assumed and no \PSg decay is considered, corresponding to a sufficiently large scalar mass scale~\cite{Giudice:2004tc,ARKANIHAMED20053, Hewett_2004, Kilian_2005}.

In addition, signals are available with a quasi-stable \stopq, the supersymmetric partner of the top quark, with masses ranging from 400 to 2600 GeV. These signals can arise, for example, in R-parity violating SUSY scenarios~\cite{Csaki_2014,Barbier_2005}.
For purposes of this study, we use long-lived \PSg and \stopq hadronized into \rhadrons~\cite{Fairbairn:2006gg,Bauer:2009cc}.
The fraction, $f$, of produced \PSg  hadronizing into a \PSg-gluon state ($R$-gluinoball) is a free parameter of the hadronization model and affects the fraction of \rhadrons\ that are neutral at production. 
The results of this search are interpreted with $f=0.1$ to allow comparisons with previous versions of this analysis.
The \rhadron's interactions with matter follow the model in Refs.~\cite{Kraan:2004tz, Mackeprang:2006gx}.

The second type of signal consists of HSCPs that behave like leptons. 
Supersymmetric models featuring gauge mediated SUSY breaking (GMSB)~\cite{Giudice:1998bp} or general gauge mediation~\cite{Meade_2009} commonly result in long-lived charged sleptons whose decay depends on the SUSY-breaking scale. 
A long-lived tau slepton (\stau) would traverse the detector in a similar way to a slow muon.
Production of \stau particles at the LHC can occur either directly or in the decays of heavier supersymmetric particles (mainly \squark and \PSg).
The GMSB model is explored using the  Snowmass points and slopes line 7 (SPS7)~\cite{Allanach:2002nj}, which has the \stau as the next-to-lightest SUSY particle.
The particle mass spectrum and the decay table are calculated with the program ISASUGRA~\cite{Paige:2003mg} v7.69.
The GMSB parameter $\Lambda$ is the effective supersymmetry breaking scale and is varied with fixed parameters for the number of messenger particles, 
${N}_{\text{mes}} = 3$,  and their mass scale ${M}_{\text{mes}}/ \Lambda = 2$. 
The ratio of the vacuum expectation values of the Higgs doublets, $\tan \beta$, is set to 10, while the sign of the higgsino mass term, $\mu$, is assumed to be positive. 
We set the scale factor of the gravitino coupling, $\Cgrav = 10^4$. 
The large value of \Cgrav\ results in a long-lived 
\stau, while $\Lambda=31$--160\TeV gives a \stau  mass of 200--1600\GeV. 
Two \stau samples are generated for each set of SUSY parameters: one with all processes (labeled ``GMSB SPS7, \stau NLSP") and one with only direct, DY pair production (labeled ``DY \stau").
For direct pair production, different possible \stauR and \stauL mixing assumptions are considered.

The third type of signal is based on modified Drell--Yan (DY) production of long-lived lepton-like fermions (\PGtpr).  
In this scenario, new massive spin-1/2 particles have arbitrary electric charge but are neutral under $SU(3)_{C}$ and $SU(2)_{L}$, and therefore couple only to the photon and the \PZ boson. 
Simulated samples for \PGtpr fermions are produced with masses ranging from 200 to 2600\GeV and for electric charges ${Q}=1e$ and $2e$.

The last type of signal considered originates from the proposal in Ref.~\cite{Giudice_2022} as a possible explanation for the ATLAS excess.
The proposed signature is considered using a minimal simplified model with the production of a heavy \PZpr boson decaying into two doubly charged \PGtpr fermions. 
Limits are set on the cross section for this simplified model, and no assumption is made about the coupling structure of the \PZpr boson. 
The \ZPrimepsi and \ZPrimeSSM production cross sections are used as benchmarks in the interpretation. 
The mass of the \PZpr boson (\PGtpr fermion) is considered in the range 3000--7000 (200--1400)\GeV.

The response of the CMS detector is simulated with the \GEANTfour~\cite{GEANT4} package. 
The simulated events are reconstructed using the same algorithms that are used for data.
In all signal samples, simulated minimum bias events are overlaid with the primary collision to produce the effect of additional interactions in the same 
or adjacent beam crossing (pileup)~\cite{simulation}.

Although the background estimate performed in this search is purely data-driven, 
a set of MC samples for \PW+ jets, top quark pair  (\ttbar), and quantum chromodynamics (QCD) multijet production has been used to
guide the selection criteria and
understand qualitatively the nature of surviving background events.

\section{Ionization observables}
\label{sec:variables}

Ionization observables are based on the cluster charges deposited 
in the silicon pixel or strip detector~\cite{Chatrchyan_trk10,Chatrchyan:1704291}.
Dedicated corrections and filters are applied.
In the strip tracker, the collected charges are corrected first by a gain calibration to stabilize the values
over time by adjusting the most probable value of the charge to the energy loss by an MIP. A special correction is applied to estimate the charge deposited by the particle in case of saturation of the electronics.
To reduce the rate of contamination from clusters with large energy deposition due to overlapping tracks
that could occur at high pileup or in boosted hadron decay,
a cleaning procedure is applied to remove clusters in the silicon strip tracker that are not consistent with the passage of a single charged particle.
Strip clusters presenting a structure with multiple peaks in the 
distribution of the charge deposits along the strips are rejected to be consistent with the hypothesis of a single-particle deposit.
Furthermore, it is required that the clusters are not located on the edge of a module in order to avoid biased measurements.
These two last criteria are referred to as the ``cluster cleaning". 
In the pixel detector, clusters are included in the ionization observable only if they are not positioned along the edges of pixel sensors, 
do not span two read-out chips, and do not correspond to known defective pixel channels.

A \dedx discriminant, \gstrip, 
is used to distinguish SM particles from HSCP candidates~\cite{Khachatryan:2011ts,Chatrchyan:2012sp,Khachatryan:2016sfv,EXO-12-026}.
Only silicon strip tracker clusters are considered in 
the discriminant, which is given by Eq.~(\ref{eq:gstrips}), 
\begin{equation}
	\gstrip = \frac{3}{N} \, \left(
   \frac{1}{12N} + \sum_{j=1}^N
   \left[
   P_j \, \left( P_j - \frac{2j-1}{2N} \right)^2 \right] \right),
   \label{eq:gstrips}
\end{equation}
where $N$ is the number of measurements in the silicon strip detectors, $P_j$ is the probability
for an MIP with the observed path length in the detector to produce a charge smaller or equal to that of the $j$-th measurement 
in the detector, and the sum is over the track measurements ordered in terms of increasing $P_j$.
The probabilities are determined using three-dimensional templates of path-length-normalized charge distributions per silicon strip detector module geometry and
per path length, and as a function of the pileup. 
These templates are derived from data passing the same selection criteria as described in Section~\ref{sec:presel} but within the momentum range of 20--48\GeV,
for nine time periods covering the 2017--2018 data taking. The range was chosen to ensure that the tracks used to create the \gstrip\ templates are representative of background tracks in this analysis.

A second  \dedx discriminant, \fpix, which is introduced in this paper, relies only on the pixel detector information and is given by Eq.~(\ref{eq:fpixel}),
\begin{equation}
	\fpix = 1 -  \prod_{j=1}^n P'_j \sum_{k=0}^{n-1} \frac{[-\ln( \prod_{j=1}^n P'_j)]^k}{k!},
	\label{eq:fpixel}
\end{equation}
where $n$ is the number of measurements in the silicon pixel detector, excluding the first barrel layer,
which was unstable during the considered data-taking period,
and $P'_j$ is the angle dependent Vavilov probability that an MIP would produce a charge larger than the $j$-th
measurement predicted by a detailed simulation (called \textsc{PixelAV}~\cite{Chiochia:2004qh}), which includes effects 
from bias voltage changes, gain calibration changes, temperature, and irradiation dose. 
The use of Vavilov probabilities correctly handles distribution shape effects in long clusters. Since a parameterized function is used to calculate the probabilities, 
the method does not require large calibration samples in the tails of the distribution but can estimate the probabilities of outlier events reliably.
The \textsc{PixelAV} electric field maps are taken from ISE TCAD~\cite{tcad} simulations of a pixel cell that have
been tuned to agree with in situ measurements of the Lorentz drift vs. substrate depth.  The carrier
trapping rates are determined from in situ charge versus depth measurements.  The simulation
also includes carrier diffusion and charge induction on the sensors. The electronics simulation 
includes noise, linearity, thresholds, and miscalibration. The Vavilov parameter templates have been produced 
25 times over the 2017--2018 period, corresponding to an update about every 5\fbinv. 
Each set of parameters is checked against low-\pt data taken during the same calibration period.

Finally, the most probable value of \dedx for a track is estimated using the squared harmonic mean, as defined in Eq.~(\ref{eq:Ih})~\cite{Chatrchyan_trk10}, 
which suppresses the tails of the Landau distribution:
\begin{equation}
	 \Ih = \biggl( \frac{1}{N} \sum_{j=1}^{N} \left(\dd{E}{x_j}\right)^{-2} \biggr)^{-1/2},
	 \label{eq:Ih}
\end{equation}
where \ddinline{E}{x_j} is the charge per unit path length of the $j$th track measurement in the sensitive part of the silicon detector, in units of\unit{\MeV/cm},
and the summation is over the $N$ measurements in the silicon strip detector.

The mass ($m$) of a candidate particle can be calculated~\cite{EXO-12-026} from its momentum ($p$) and  \Ih, 
based on a relationship approximating the Bethe--Bloch formula in the nonrelativistic regime as in Eq.~(\ref{eq:MassFromHarmonicEstimator}), 
\begin{equation}
	\Ih = K\frac{m^2}{p^2}+C,
 	\label{eq:MassFromHarmonicEstimator}
\end{equation}
where the empirical parameters $K$ and $C$ are determined using a sample of low-momentum particles composed of protons, kaons, and pions.  

The values of $K$ and $C$ for the 2017 and 2018 data samples are listed in Table~\ref{tab:k_and_c_values}. The uncertainties associated with the measurements take into account the variation of the data with time and small differences observed between different MC samples.
By inverting Eq.~(\ref{eq:MassFromHarmonicEstimator}), the mass is estimated assuming a $Q=1e$ particle.

\begin{table}[ht!]
	 \centering
  \topcaption{Summary of $K$ and $C$ values for data, and simulation.}
  \label{tab:k_and_c_values}
  \begin{tabular}{ccccc }  
   Parameter & \multicolumn{2}{c}{Data} & \multicolumn{2}{c}{Simulation} \\ 
   (\MeVns/\cm) &    2017  &  2018  & 2017 & 2018  \\ \hline
$K$  &  2.54 $\pm$  0.05   &  2.55 $\pm$ 0.05 &  2.50 $\pm$ 0.05 & 2.49 $\pm$ 0.05 \\
$C$  &  3.14 $\pm$  0.02   &  3.14 $\pm$ 0.02 &  3.18 $\pm$ 0.02 & 3.18 $\pm$ 0.02 \\ 
  \end{tabular}
\end{table}

\section{Event selection}
 \label{sec:presel}

All events are selected online by a trigger requiring the reconstruction of a muon with transverse momentum $\pt > 50\GeV$.
An offline reconstructed muon passing ``tight" identification criteria~\cite{CMS:2018rym} must match the online reconstructed muon that triggered the event.
The muon trigger is more efficient for lepton-like HSCPs  (with the
efficiency varying between 0.7 for a \stau mass of 1.2\TeV and 0.9 for a \stau mass of 300\GeV) 
than for \rhadron\ HSCPs (with efficiency varying between
0.074 for a \PSg mass of 2.6\TeV and 0.19 for a \PSg mass of 1\TeV). 
The mass dependence of the trigger efficiency is understood to arise from a $\beta$ dependence: 
very slow particles ($\beta<0.5$) reach the muon detectors later than the time window of the trigger.
The reduced efficiency of the \rhadron\ HSCP compared to the \stau scenario arises because the \rhadron\
will sometimes, with a probability on the order of 60\%, change charge when interacting with matter, leaving no signal in the muon chambers.
In the case of the charge-suppressed \rhadron\ model (in which only the flip from charged to neutral \rhadron\ is allowed, not the reverse one),
the efficiency is almost zero; this particular model is not considered for interpretation in this paper.
 
The HSCP candidates are required to be associated with a track reconstructed from hits in the silicon detectors.  
All events are required to have at least one HSCP candidate with $\pt > 55\GeV$.
The pseudorapidity range is restricted to $\abs{\eta}< 1.0$ to ensure that  
the geometrical properties of the silicon sensor modules are uniform 
and to improve the signal to background ratio.
To remove mismeasurements that could bias both the momentum and the ionization,
a requirement on the  \pt\ uncertainty divided by $\pt^2$ is applied: $\sigptoverptsq < 0.0008\GeV^{-1}$,
which does not impact the signal efficiency for  HSCP candidates at large mass.
After cluster cleaning described in Section~\ref{sec:variables}, the candidate must have at least 10 measurements in the silicon tracker.
Furthermore, to ensure good ionization measurement for the track, the HSCP candidates are required to have at least two measurements in the silicon pixel detector, a fraction of valid hits higher than 80\%, pass quality (high purity) requirements, and possess a track fit value with $\chi^2/\mathrm{dof} < 5$. 
The details of these tracking criteria are explained in Ref.~\cite{Chatrchyan:1704291}.
Since the \fpix\ variable is also sensitive to charges that are less than the expectation from an MIP, 
it can be used to remove truncated or broken clusters. Clusters can become truncated at module edges
and dead regions of the detector. Dynamic readout inefficiencies can cause large clusters to break
into smaller parts. Requiring \fpix\ to be larger than 0.3 removes tracks with truncated and broken clusters.
The magnitudes of the longitudinal and transverse impact parameters with respect to the primary vertex, \dz\ and \dxy, are required to be less than 0.1\unit{cm} and 0.02\unit{cm}, respectively, to reduce the contribution from pileup tracks and ensure that the track is originating from the hard collision.

Candidates must also pass isolation requirements.
To remove the contribution to the background from events composed primarily of jets produced through the strong interaction, 
a relative mini-isolation variable \Irel\ based on PF objects is required to be less than 0.02, with $I_{\text{PF}}^{\text{rel}}$ defined as in Eq.~(\ref{eq:Irel}), 
\begin{equation}
	\Irel = \frac{(\sum_{\Delta R<\DeltaRmini} \pt^{\text{PF}})-\pthscp} {\pthscp},
	\label{eq:Irel}
\end{equation}
where \pthscp\ indicates the \pt\ of the HSCP candidate.
The sum runs over all PF charged and neutral hadrons, photons, electrons, and muons, and takes pileup into consideration as well, within 
a \pt-dependent cone selection~\cite{Rehermann_2011} as specified in Eq.~(\ref{eq:DeltaRmini}), 
\begin{equation}
	\DeltaRmini = \begin{cases}
         0.2, & \text{ $\pthscp < 50\GeV$} \\
		10 \GeV/\pthscp, & \text{ $50\GeV < \pthscp < 200\GeV$ } \\
		0.05, & \text{ $\pthscp > 200\GeV$}.\\
    \end{cases}
    \label{eq:DeltaRmini}
\end{equation}
In particular, the inclusion of muons in the  \Irel\ definition removes the case of two collimated muons, that could come for example 
from boosted resonances such as \JPsi, which could lead to a large \dedx deposit if the two muon tracks overlap. 
An absolute track-based mini-isolation requirement $\Itrk <  15\GeV$ is also applied, with \Itrk\ defined
as in Eq.~(\ref{eq:Itrk}), 
\begin{equation}
	\Itrk = \biggl( \sum_{\Delta R<\DeltaRmini}\pt^{\text{trk}} \biggr) -\pthscp,
	\label{eq:Itrk}
\end{equation}
in which the sum runs over all tracks associated with the primary vertex that are within the mini-isolation cone.
This absolute criterion on the \Itrk\ isolation complements the relative \Irel\ selection by
 rejecting HSCP candidates with remnant activity close by. Such activity could induce a large
 \dedx value for the HSCP track if another track (from PU or other sources) crosses the module at the same place as the HSCP track. 
A last condition applied in the candidate selection, aimed at removing electrons, is the requirement
$E/p < 0.3$, where $E$ is the energy of the associated PF candidate measured by the calorimeters.

In the case when several HSCP candidates per event are present, the track with the highest \ih\ is used.
Applying a looser ionization selection and categorizing events with exactly one or two candidates does not lead to any significant improvement.

The selection requirements are summarized in Table~\ref{tab:cutflow} with the corresponding
cumulative efficiencies for the data and two signal hypotheses. In data, the trigger efficiency represents the selection from the primary data set with muon triggers to the specific triggers used in this analysis.

\begin{table}[ht]
	\centering
	\topcaption{Cumulative selection efficiency for the data and for two signal hypotheses.}
  	\label{tab:cutflow}
	\begin{tabular}{lccc}
Selection criteria & Data & \PSg (1.8\TeV) & DY \stau (557\GeV) \\
\hline
All events &    1  &   1  &   1  \\
Trigger &  0.15  & 0.11  & 0.86  \\
$\pt > 55\GeV$ &  0.11  & 0.11  & 0.86  \\
$\abs{\eta} < 1$ &  0.059  & 0.074  & 0.64  \\
\# of valid pixel hits in L2-L4 $\ge 2$  &  0.056  & 0.071  & 0.62  \\
Fraction of valid hits  $> 0.8$  &  0.052  & 0.069  & 0.62  \\
\# of \dedx measurements $\ge 10$ &  0.052  & 0.069  & 0.62  \\
High-purity track &  0.052  & 0.069  & 0.62  \\
Track $\chi^2$/dof $< 5$  &  0.052  & 0.069  & 0.62  \\
$\dz < 0.1$\unit{cm}  &  0.052  & 0.069  & 0.62  \\
$\dxy < 0.02$\unit{cm} &  0.048  & 0.069  & 0.62  \\
$\Irel < 0.02 $ &  0.014  & 0.065  & 0.61  \\
$\Itrk  < 15\GeV$ &  0.014  & 0.065  & 0.61  \\
PF $E/p < 0.3$ &  0.014  & 0.064  & 0.61  \\
\sigptoverptsq $< 0.0008$ &  0.014  & 0.064  & 0.61  \\
\fpix $> 0.3$ &  0.011  & 0.064  & 0.60  \\
	\end{tabular}
\end{table}

\section{Background prediction methods and dedicated signal and validation regions}
\label{sec:bkgpred}

Two different data-based estimation methods have been developed to predict the background yields, allowing cross checks to be made: the ionization method and the mass method.

\subsection{The ionization method}

\begin{figure}[ht!]
   \centering
       \includegraphics[width=.42\textwidth]{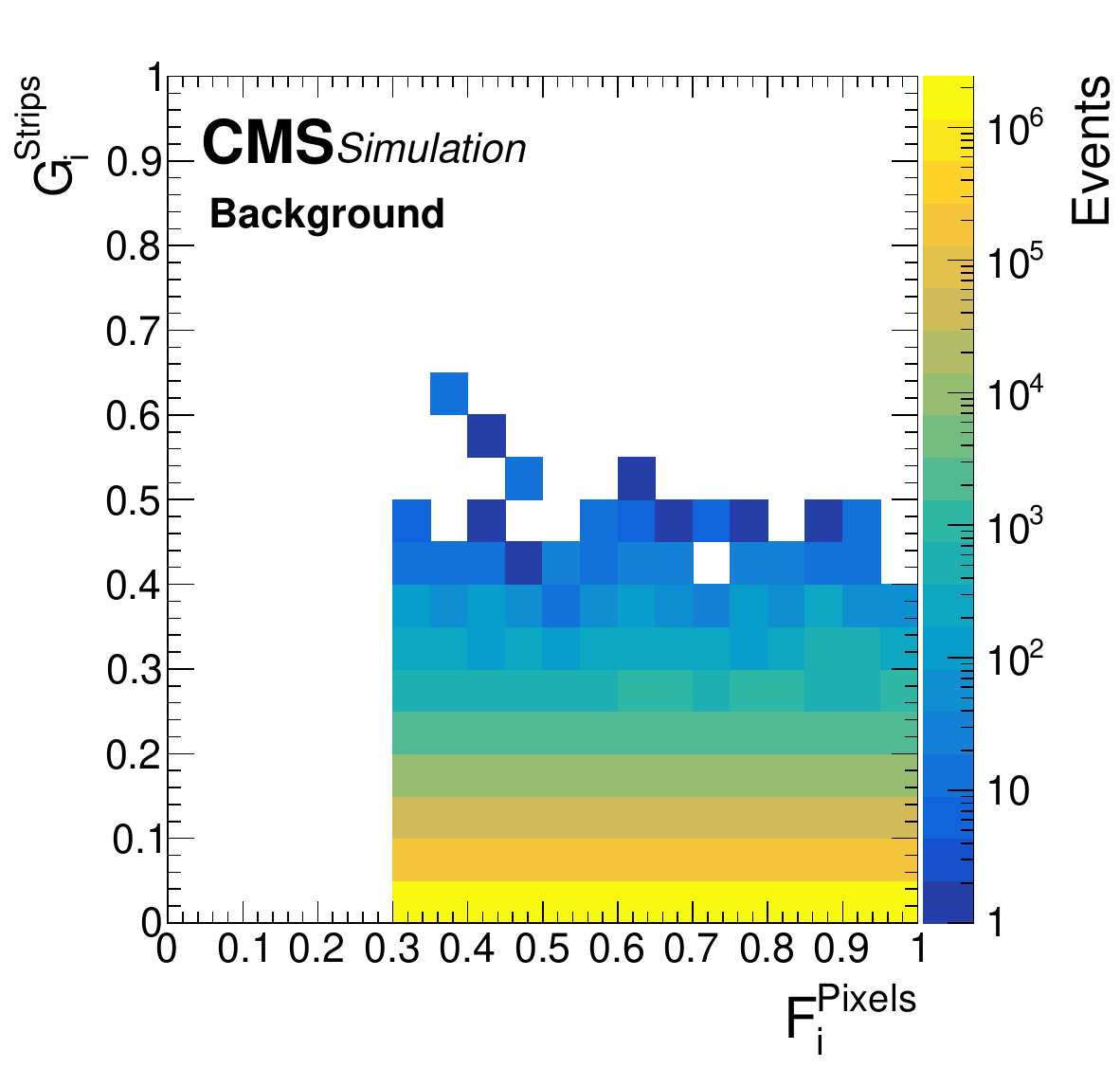}
       \includegraphics[width=.42\textwidth]{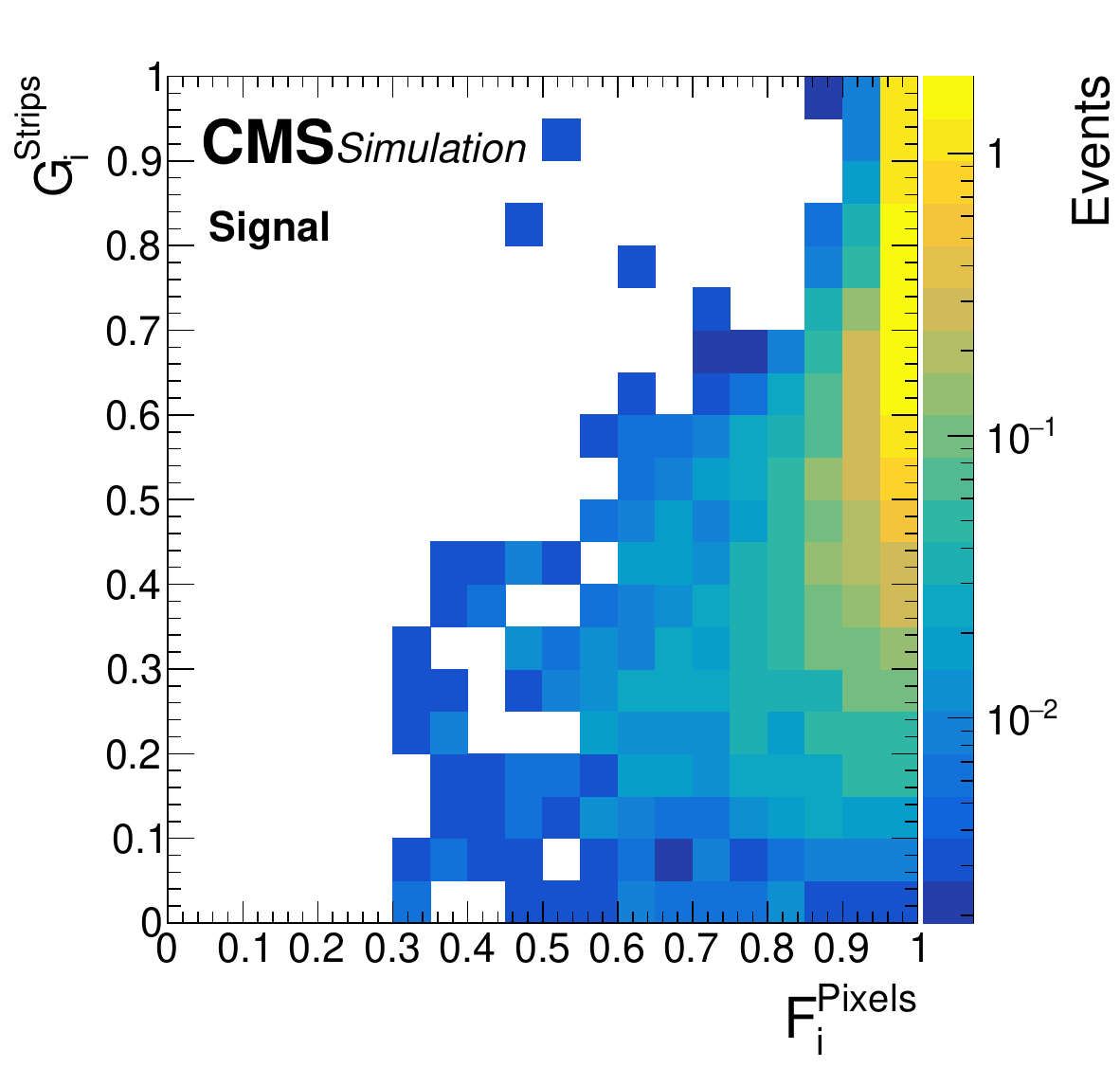}
	\caption{The simulated \fpix\ vs. \gstrip\ distribution for the SM background (left), and an 1800\GeV \PSg R-hadron (right),
	for events that pass 
	the selection criteria listed in Table~\ref{tab:cutflow}.}
   \label{fig:GiVsFi}
\end{figure}

The ionization method is based on two uncorrelated ionization variables \fpix\ and \gstrip, 
which pertain to independent detectors.
Figure~\ref{fig:GiVsFi} shows the \fpix\ vs \gstrip\ distribution for the SM background simulation (left) and the simulation of 1800\GeV mass \PSg R-hadron pair production (right) for 
candidates passing the event selection criteria described in Section~\ref{sec:presel}. 
The signal peaks in the high-\fpix\ and high-\gstrip\ corner, while for the background the distribution in \fpix\ is flat,
and \gstrip\ follows a falling exponential-like shape.

Additional criteria on variables describing the ionization are applied to improve the discrimination
of signal and background.
The  signal region (SR) of the ionization approach is defined by $\fpix >0.9$, $\gstrip > 0.25$, and $\pt > 200\GeV$. 
The cut on the transverse momentum strongly reduces the SM contribution while keeping the signal efficiency high.
This \pt threshold is optimized with respect to the expected cross section upper limits.
The optimization uses the full mass ranges of both the \PSg \rhadrons\ and \stau samples.

A fit is performed on the entire \gstrip\ distribution ($0 \le \gstrip \le 1$)
in two regions simultaneously: the FAIL region defined by $0.3< \fpix <0.9$ and the PASS region 
defined by $\fpix >0.9$. 

The background template used in the PASS region is estimated as the observed data distribution in the FAIL region, multiplied by a transfer function, through the relation in Eq.~(\ref{eq:Rpf}),
 \begin{equation}
	 \NPASS = R_{\text{P/F}}(j) \, \NFAIL ,
	 \label{eq:Rpf}
\end{equation}
where \NPASS\ is the background yield in $j$-th bin of the  \gstrip\ distribution in the PASS region,
\NFAIL\ is a freely floating parameter that determines the background yield in the $j$-th bin, which is constrained to values close to the observed data yield during the fit, and 
 \RPFj\ is the value of the pass-to-fail ratio for bin $j$. The \RPF\ ratio is modeled as a constant across the bins $j$,
 whose value is determined during the fit. Higher-order polynomials were tested using Fisher's $F$-test~\cite{WileyChi2,BAKER1984437} and provided 
 results as good as the constant function, 
 but not significantly better.
The binning of the \gstrip\ distribution was chosen so the last bin contains $\mathcal{O}(1)$ event in the FAIL region for the data.

The background
uncertainty has two sources: the statistical uncertainty in the background yield in the FAIL region (which is dominant in the tail of
the \gstrip\ distribution) and the statistical uncertainty in \RPF.

\begin{figure}[ht!]
   \centering
       \includegraphics[width=.42\textwidth]{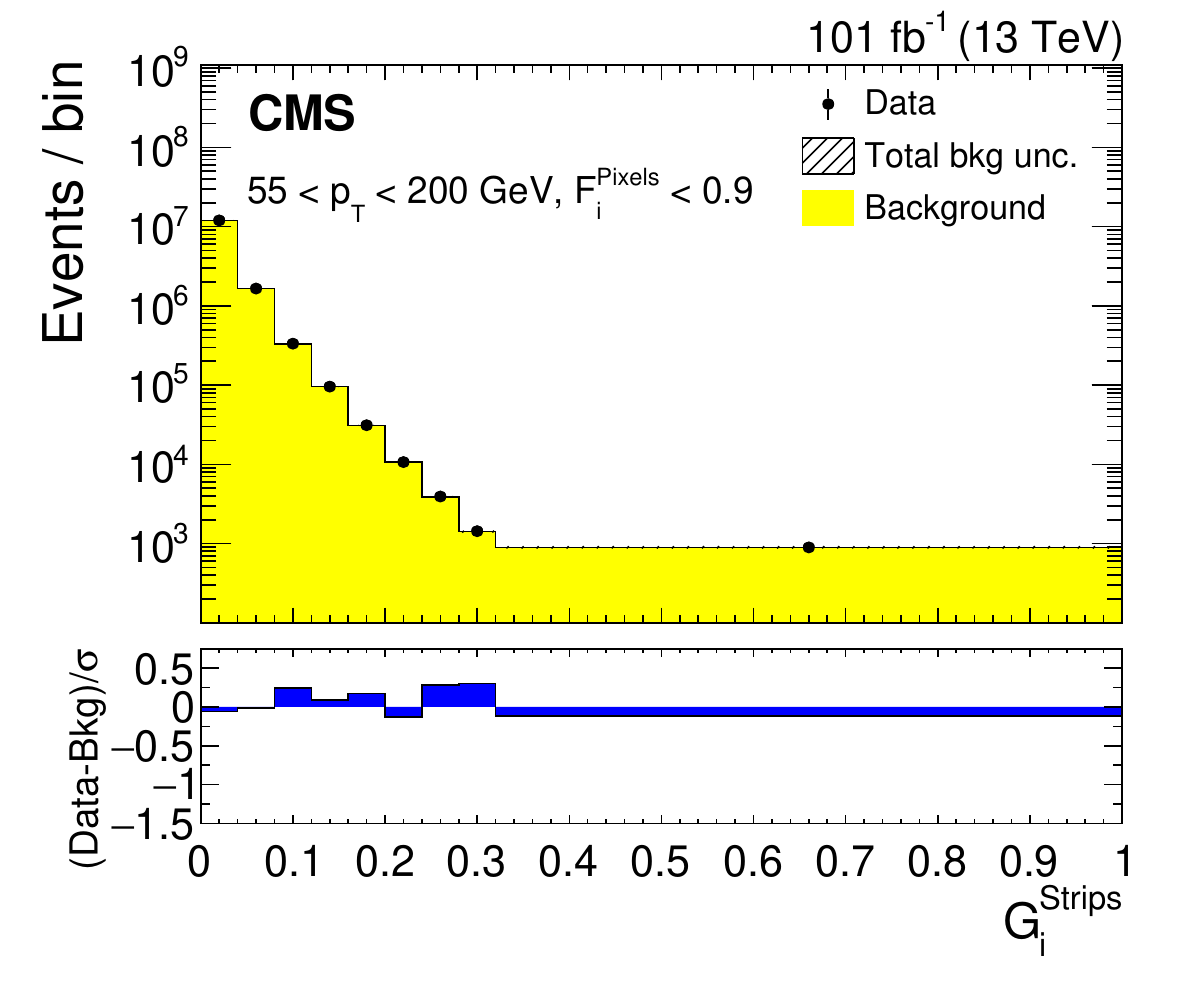}
       \includegraphics[width=.42\textwidth]{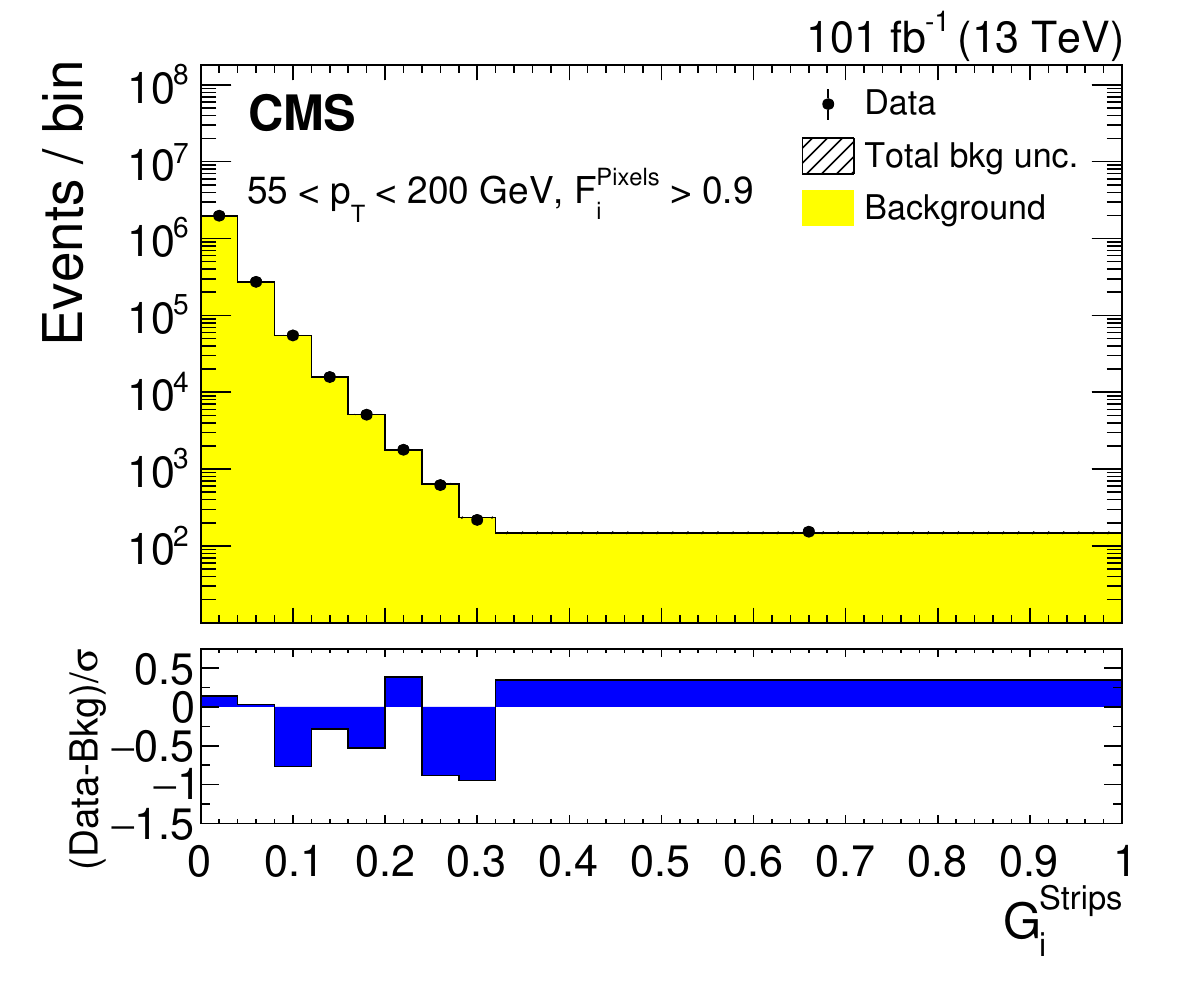}
        \caption{The \gstrip\ distribution in the FAIL (left) and PASS (right) regions for events passing the event selection and with	$55  < \pt <  200\GeV$.
        The data are represented by black markers.
        The background predicted by the ionization method  is shown in yellow, with the shaded area indicating the scarcely visible background uncertainty.
	The lower panel displays the pulls, defined as the difference between the data and the estimated background, divided by the associated uncertainty.}
   \label{fig:valid_check}
\end{figure}

The method is tested successfully with MC simulations and data. The validation with data is performed in two validation regions defined
by lower \pt\ ranges: $50 < \pt < 55\GeV$ and $55 < \pt <  200\GeV$. 
Figure~\ref{fig:valid_check} shows  the \gstrip\ distribution in the FAIL and PASS regions for candidates with $55  < \pt <  200\GeV$. The histograms associated with the 557\GeV \stau and 1800\GeV \PSg signals are not visible on these plots, their contribution being negligible in this region.

\subsection{The mass method}

The mass method is based on searching for a new resonance in the mass spectrum of the HSCP candidates using the ionization
variable  \Ih\ and  momentum $p$, calculated by inverting Eq.~(\ref{eq:MassFromHarmonicEstimator}) under the assumption $Q=1e$ for all signals considered in this analysis.
The implementation used here improves on the previously published analysis~\cite{Khachatryan:2016sfv} in the background estimation procedure as well as in the tightened selections.

For the mass method, further requirements have been applied to the event selection in addition to those described in Table~\ref{tab:cutflow}: 
$\Ih  > C$ (with $C$ introduced in Eq.~(\ref{eq:MassFromHarmonicEstimator})), $\Iconetrk < 15\GeV$, where the 
\Itrk\ isolation variable is computed in a fixed cone of $\Delta R$ of 0.3, and $\sigmaptoverpt <1$. The two last criteria reduce the
background contribution in the high mass region.

The mass method relies on the lack of correlation between momentum and ionization measurements for background candidates, and divides the events into four regions:
\begin{itemize}
\item the control region A defined as $\gstrip < 0.018$ and $55 <\pt < 70\GeV$,
\item the control region B defined as $\gstrip > 0.22$ and $55 <\pt < 70\GeV$,
\item the control region C defined as $\gstrip < 0.018$ and $\pt > 70\GeV$, and
\item the signal region D is defined as $\gstrip > 0.22$ and $\pt > 70\GeV$. 
\end{itemize}	
The fraction of events with $\gstrip > 0.22$ corresponds to approximately 0.1\% of the remaining SM background,
while the fraction of events with $\gstrip < 0.018$ corresponds to about 50\%.
The normalization of the background in signal region D ($N_{bkg}(D)$) is determined using Eq.~(\ref{eq:ABCDNorm}), where $N(A)$, $N(B)$, and $N(C)$ are the yields in the control regions A, B, and C, respectively. 
\begin{linenomath}
\begin{equation}
N_{bkg}(D) =\frac{N(B)N(C)}{N(A)}
\label{eq:ABCDNorm}
\end{equation}
\end{linenomath}

The shape of the background mass spectrum in region D comes from inverting Eq.~(\ref{eq:MassFromHarmonicEstimator}) with fitted distributions of \Ih\ in the control region B 
and $p$ in the control region C. The procedure is repeated in bins of pseudorapidity
to take the correlation between \pt\ and $p$ into account.
In order to compute the statistical uncertainty associated to the prediction,
pseudoexperiments are used in each $\eta$ bin, randomizing the \Ih\ and $p$ templates used for the mass spectrum
and the yields in the control regions for the global normalization.  The average of these 200 pseudo mass-predictions
is used as the central value for the mass prediction, the root mean square as the statistical
uncertainty envelope.

Validation regions defined with smaller \gstrip\ intervals (up to $\gstrip = 0.057$) are used to test the method.
A trend is observed in the ratio of the observed mass spectrum to the 
predicted spectrum in these validation regions.
The source of this effect is understood as coming from residual effects such as small nonlinear correlations 
between \Ih\ and $p$, which are not taken into account by the method.
The bias of the method is estimated by performing a fit to this ratio
at low mass values $m<300$ \GeV, and used to  correct the predicted mass spectrum at high mass.

Various systematic uncertainties related to the binning and the fits
for the background prediction are evaluated. 
The largest uncertainty is associated with the bias correction, and is computed by not applying the bias correction
or applying it twice.

The signal region for the mass approach is defined as $\gstrip > 0.22$, 
$\pt> 70\GeV$, and $m>300\GeV$. For each signal hypothesis,
signal and background yields are extracted in a specific mass window defined
by [$\mu -\sigma$, $\mu + 2\sigma$] where $\mu$ is the  mean of the reconstructed signal mass spectrum and
$\sigma$ its standard deviation. This asymmetric mass window is chosen to improve the sensitivity in the presence of a steeply falling background.

The method is extensively validated  with data in validation regions at lower \gstrip\ values, as well as with MC simulations. 

\subsection{Comparison of the two methods}

The ionization approach selects high-ionization candidates while the mass
approach selects candidates in a mass window, 
which allows for the acceptance of candidates with lower ionization compared to the ionization method.
A study of signal MC samples shows that the overlap of the events in the SR selected by the two methods is about 60 to 80\% depending on the channel.
Both approaches provide similar sensitivity to the various signal processes considered here.

\section{Signal systematic uncertainties}
\label{sec:systematics}

This section describes the sources of systematic uncertainty relating  to the signal acceptance. 
The sources for the two approaches are similar.

The signal efficiencies are rescaled by the data-to-simulation correction factors for muon trigger and muon ID.
The signal efficiency uncertainty is dominated by the uncertainty in the trigger efficiency for all the models.  
The trigger efficiency  depends on the particle velocity:
HSCPs that are moving too slowly could be associated with the next bunch crossing.
Efficiency curves are derived as a function of the  generator-level $\beta$ for different bins in $\eta$,
and variations on these curves have been conservatively estimated assuming a delay of 1.5\unit{ns} in the muon chambers~\cite{CMS:2018rym}; this delay corresponds to the time resolution of the chambers.
As a systematic uncertainty, $\beta$- and $\eta$-dependent trigger weights are applied to the particle that triggered the event. 
The impact can be as large as 30\%, and is the dominant source of signal systematic uncertainty.

The stability of the \gstrip\ variable across the data-taking period is studied. The mean of this quantity varies by about 2\%. To evaluate the impact of this time variation, a smearing of the \gstrip\ variable is applied to signal samples, and smeared samples are compared to the nominal samples without smearing. This systematic uncertainty is less than~1\%.

The stability of the \fpix\ variable was studied and found to be dominated by effects due to irradiation. 
Systematic uncertainties in the  \fpix\ variable are calculated using signal samples simulated with different irradiation assumptions 
and comparing the signal efficiency in the different cases. A change of 0.5\% or less is observed when comparing the present conditions 
with the conditions at the beginning or at the end of the year. 
A conservative 0.5\% systematic uncertainty is applied on the \fpix\ variable, leading to a shape uncertainty.

The systematic uncertainties in the muon momentum resolution are estimated by varying the fit window and the number of bins when extracting the \PZ boson mass
 resolution, and amount to 10\% of the muon momentum. Systematic uncertainties in the muon trigger efficiency, the muon identification, and the muon scale factors are also 
included; they lead, however, to effects less than 1\%. Further details can be found in Ref.~\cite{CMS:2018rym}.

The integrated luminosities for the 2017 and 2018 data taking years have 1.2--2.5\% individual uncertainties~\cite{CMS-LUM-17-003,CMS-PAS-LUM-17-004,CMS-PAS-LUM-18-002}, while the overall uncertainty for the 2017--2018 period is 2.0\%.

The uncertainty in the amount of pileup is evaluated by varying the total inelastic cross section ~\cite{inelasticxsec} used to calculate the weights
 by  $\pm\,1$ standard deviation, then applying it to the signal events in order to reproduce the pileup distribution observed in data. 
The pileup uncertainty is less than~1\%.

\section{Results}
\label{sec:results}

\begin{figure}[ht!]
    \centering
        \includegraphics[width=.45\textwidth]{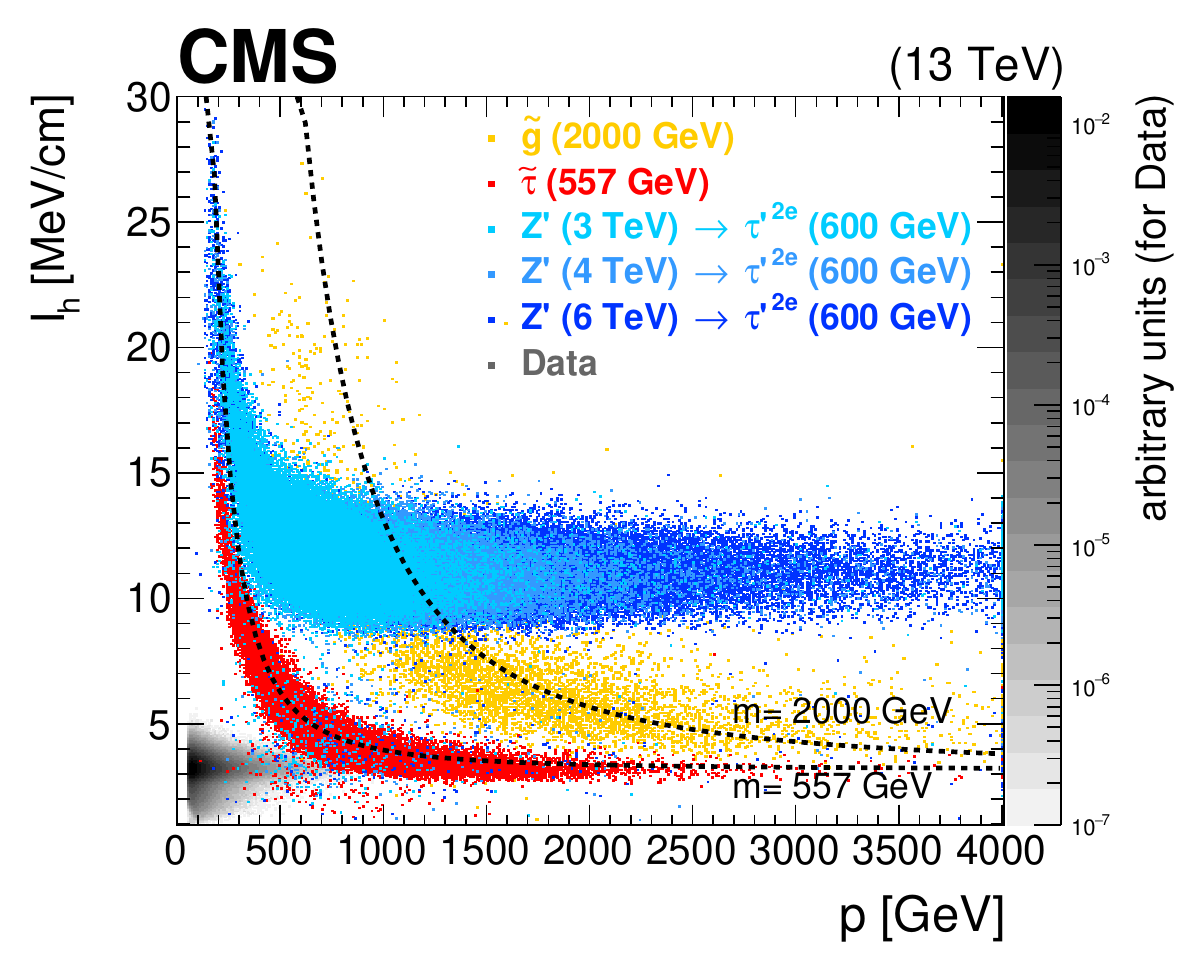}
        \includegraphics[width=.45\textwidth]{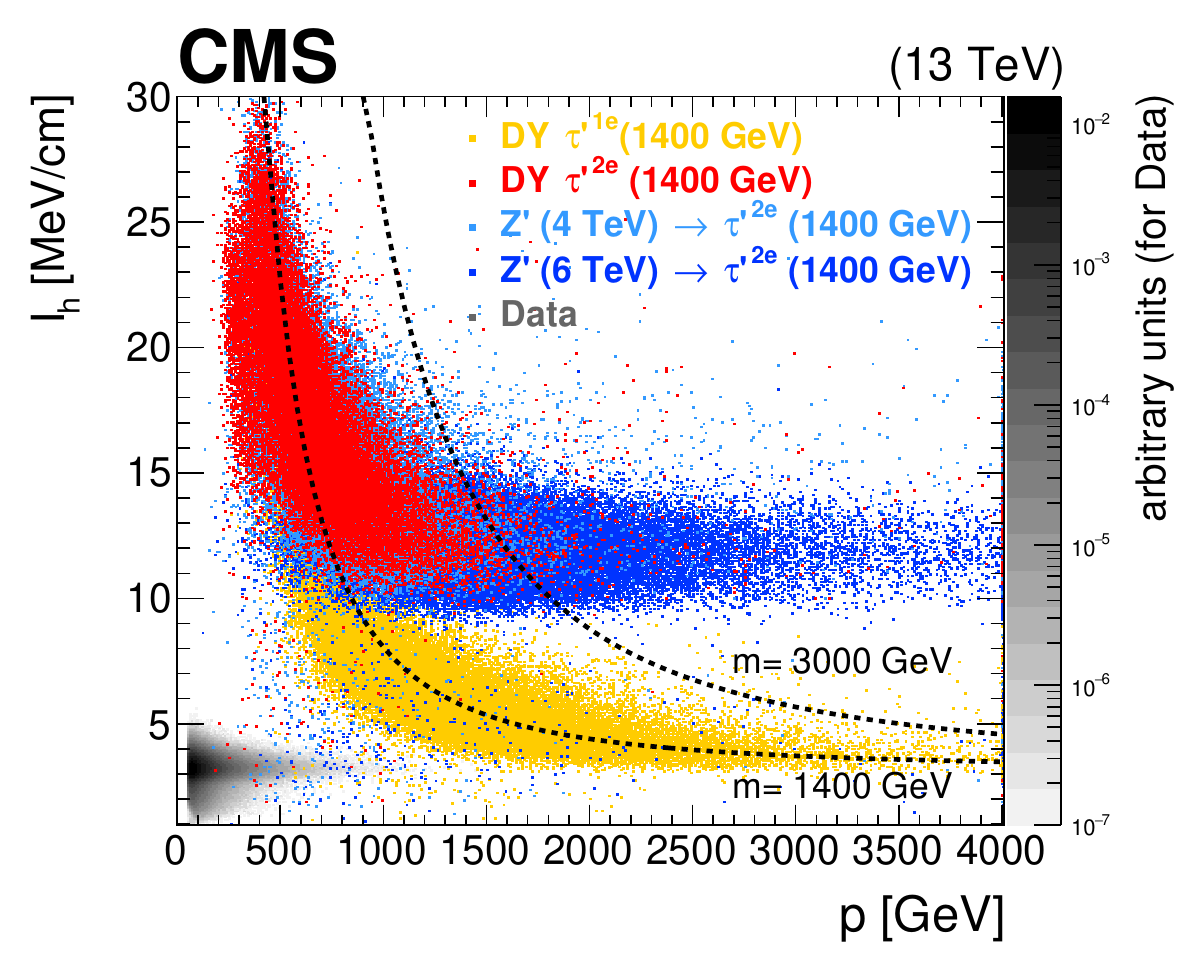}    
	 \caption{Distribution of \ih\ as a function of $p$ for the HSCP candidates passing the preselection. 
	 The colored scatter plots highlight the differences between select HSCP models with various masses and charges. 
	 For illustrative purposes, observed data are displayed using the gray density distribution, normalized to unit area. 
	 The two dashed lines based on Eq.~( \ref{eq:MassFromHarmonicEstimator})
	 correspond to a particle mass of 557 and 2000\GeV on the left, and 1400 and 3000\GeV on the right. 
	 The momentum is measured assuming a charge $1e$.}
    \label{fig:suppl_dedx}
\end{figure}

Figure~\ref{fig:suppl_dedx} shows the distribution of \ih\ as a function of $p$ for the HSCP candidates 
that pass the preselection, for various signal and mass hypotheses and for data.

\begin{figure}[ht!]
   \centering
       \includegraphics[width=.45\textwidth]{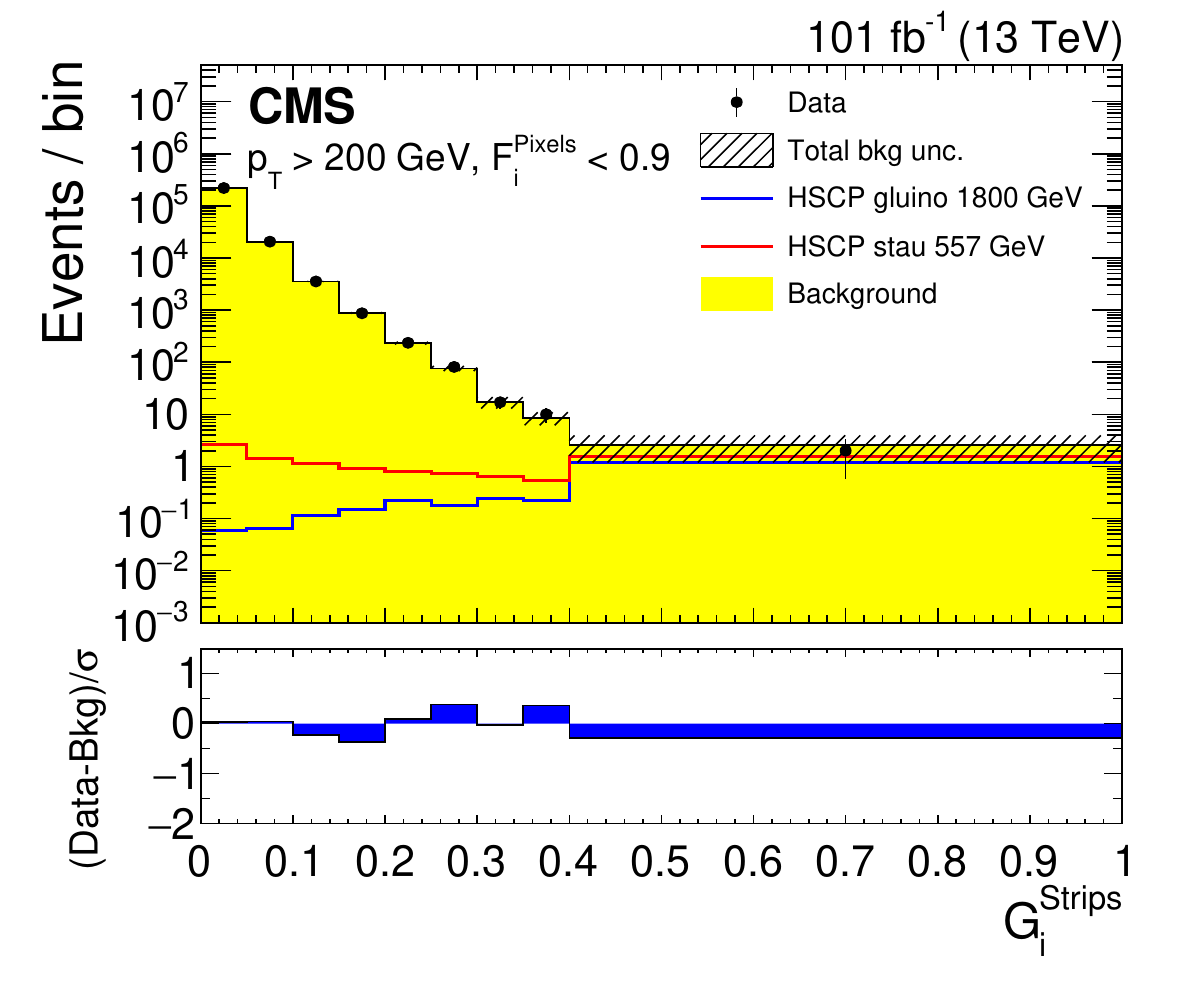}
       \includegraphics[width=.45\textwidth]{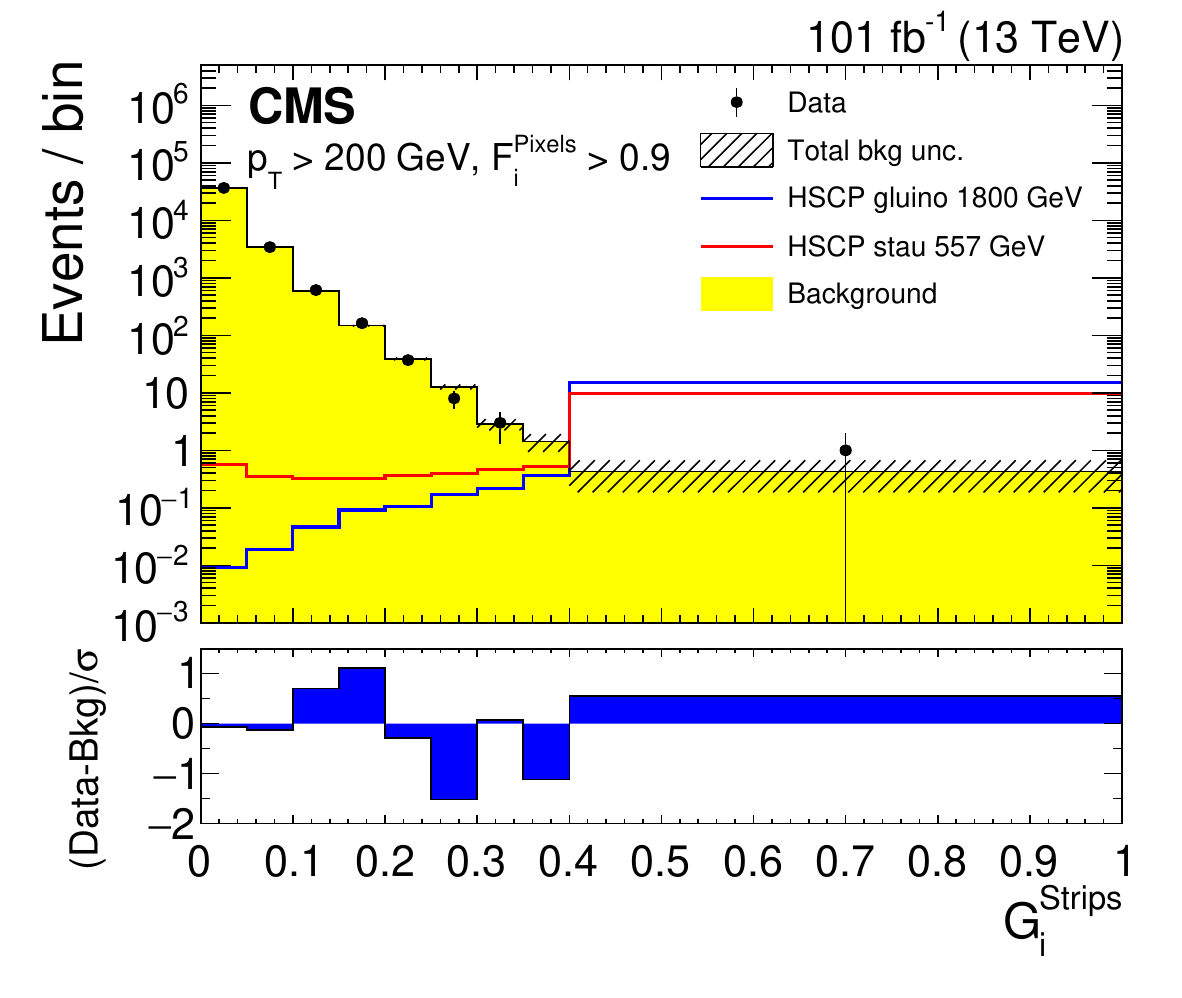}
        \caption{The \gstrip\ distribution in the FAIL (left) and PASS (right) regions for events passing the event selection and with
        $\pt > 200\GeV$. The  data are represented by black dots.
        The background predicted by the ionization method  is shown in yellow,
        with the hatched area indicating the background uncertainty.
        As examples, the blue line shows the 1800\GeV \PSg signal distribution and the red line shows the 557\GeV \stau signal distribution.
        The lower panel displays the pulls, defined as the difference between the data and the estimated background, divided by the associated uncertainty.} 
        \label{fig:Results_ioniz}
\end{figure}

Figure~\ref{fig:Results_ioniz} shows the \gstrip\ distributions for the data and the predicted background with the background-only fits performed
by the  ionization method
in the FAIL (left) and PASS (right) regions for events passing the event selection and with $\pt > 200\GeV$.
In the PASS region, one event is observed in the last bin (corresponding to $\gstrip > 0.4$) while $0.4 \pm 0.2$ events are expected. 

\begin{figure}[ht!]
   \centering
	\includegraphics[width=.6\textwidth]{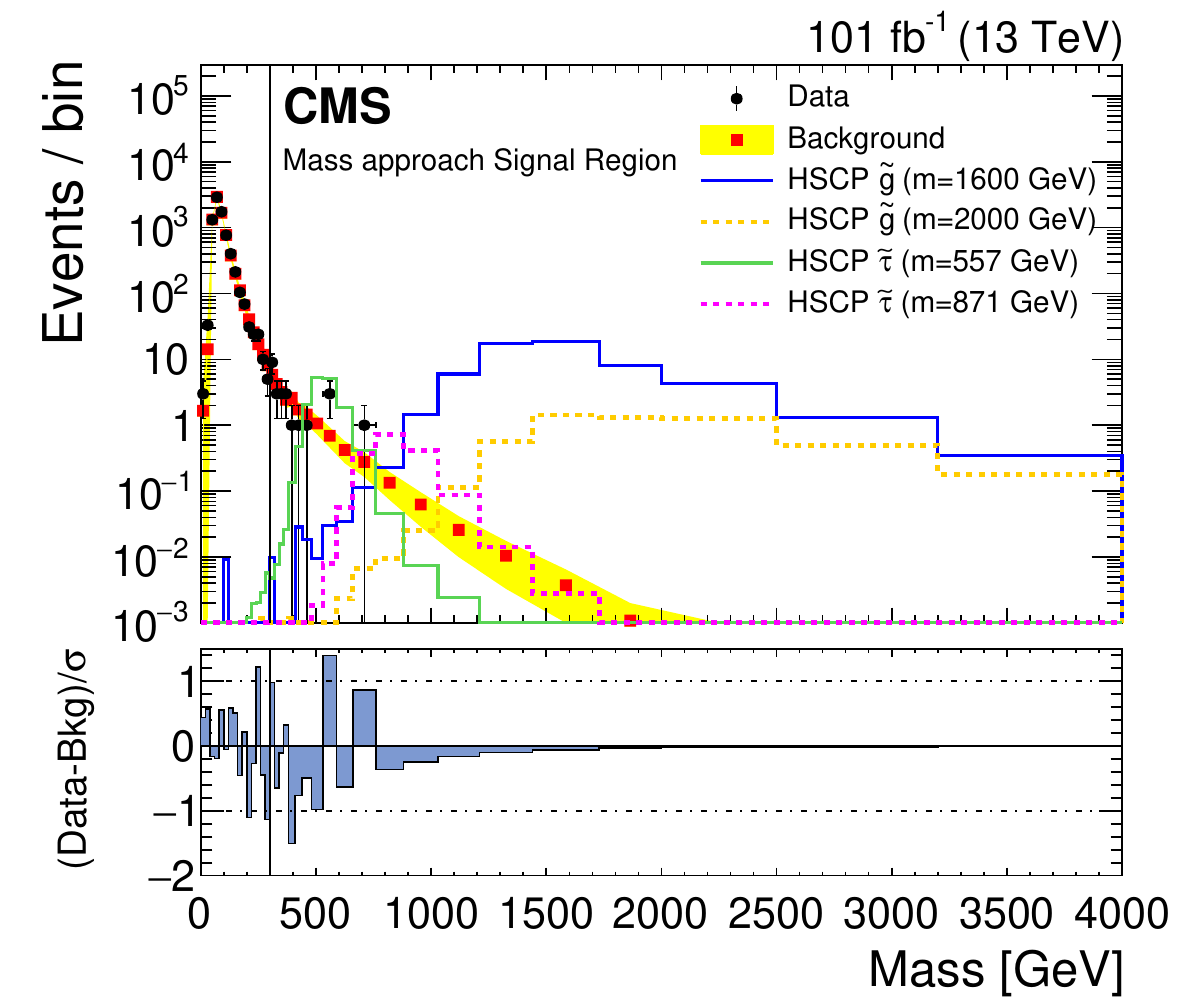}
	\caption{Mass spectrum predicted in the signal region defined by  $\gstrip > 0.22$ and  $\pt > 70\GeV$.
	The  data are represented by black dots.
        The data-driven background estimate is displayed as red markers with the yellow envelope representing 
	the quadratic sum of the statistical and the systematic uncertainties.
	Several signal scenarios are displayed. The last bin includes the overflow.
	The lower panel displays the pulls, defined as the difference between the data and the estimated background, divided by the associated uncertainty.}
	\label{fig:Results_mass}
\end{figure}

Figure~\ref{fig:Results_mass} shows the mass spectrum for the data and the predicted background for events passing the event selection, the three 
requirements on \Ih,  
\Iconetrk, and  \sigmaptoverpt\ of the mass method, and the requirements on \gstrip\ and \pt\ defining the SR: 
$\gstrip > 0.22$ and $\pt > 70\GeV$. 
The prediction describes the data well for masses larger than 300\GeV.

The two approaches show no significant excess in the data and their results are interpreted in the context of various HSCP signal models.  
The limits are calculated using the CLs criterion~\cite{Junk:1999kv, READ:JPG2002}  with the modified profiled likelihood ratio~\cite{Cowan:2010js} as test statistic.
It uses a lognormal model for nuisance parameters affecting the yields,
and shape uncertainties were parameterized as explained in Ref.~\cite{CAT-23-001}.
Given the low yields, the test statistics distributions were estimated by generating 20000  toys for the mass method and 5000 for the ionization method.
The following results have been determined using the CMS statistical analysis tool \textsc{Combine}~\cite{CAT-23-001},
which is based on the \textsc{RooFit}~\cite{Verkerke:2003ir} and \textsc{RooStats}~\cite{Moneta:2010pm} frameworks.

The  mass method provides slightly stronger expected limits for signal hypotheses with a large mass because it benefits from a very low background in this regime.
However, the background in the mass spectrum is not negligible below 600\GeV, degrading the limits at low masses.
The ionization method, which uses the same background yield for all the signal hypothesis, performs better for low mass signals.

\begin{figure}[ht!]
   \centering
	\includegraphics[width=.42\textwidth]{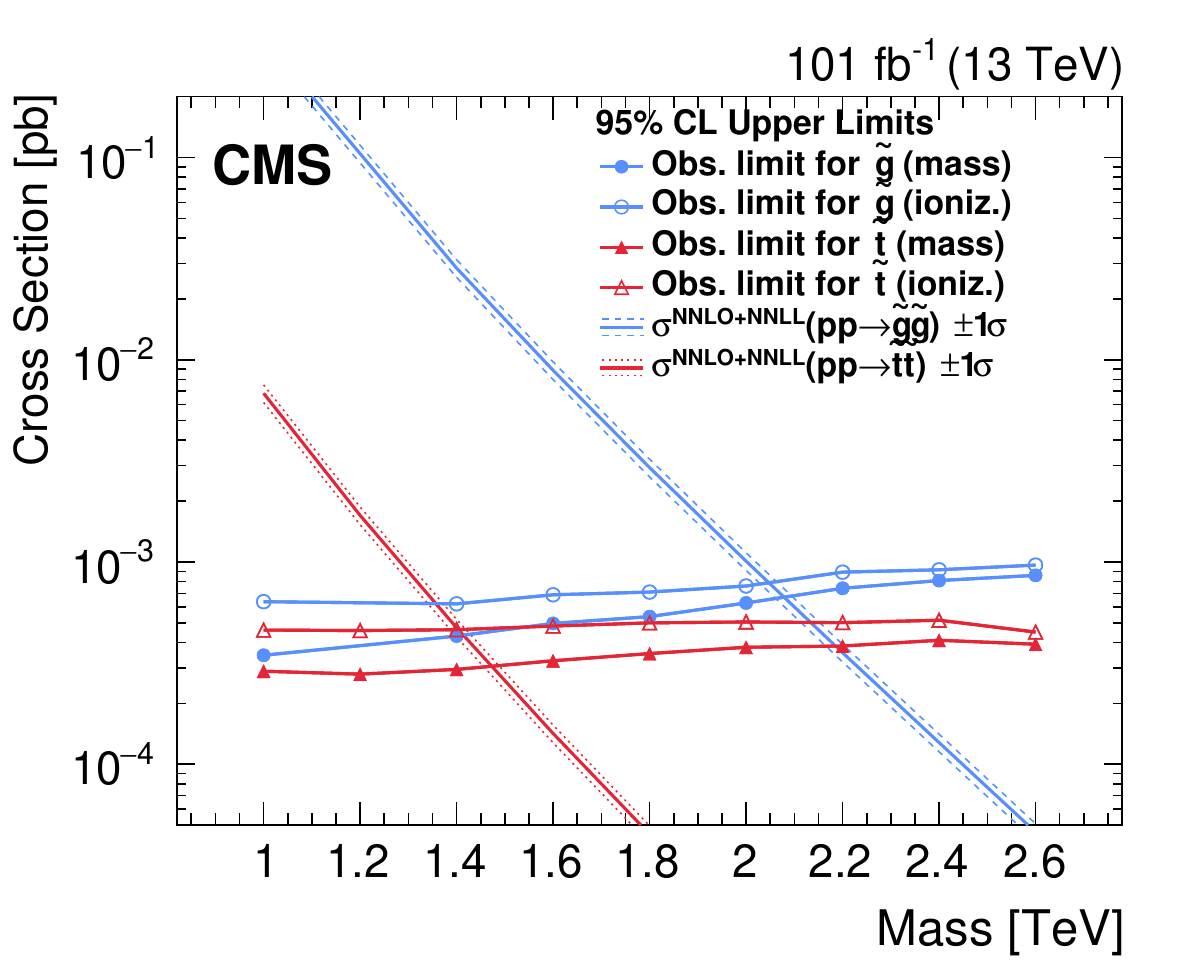}
	\includegraphics[width=.42\textwidth]{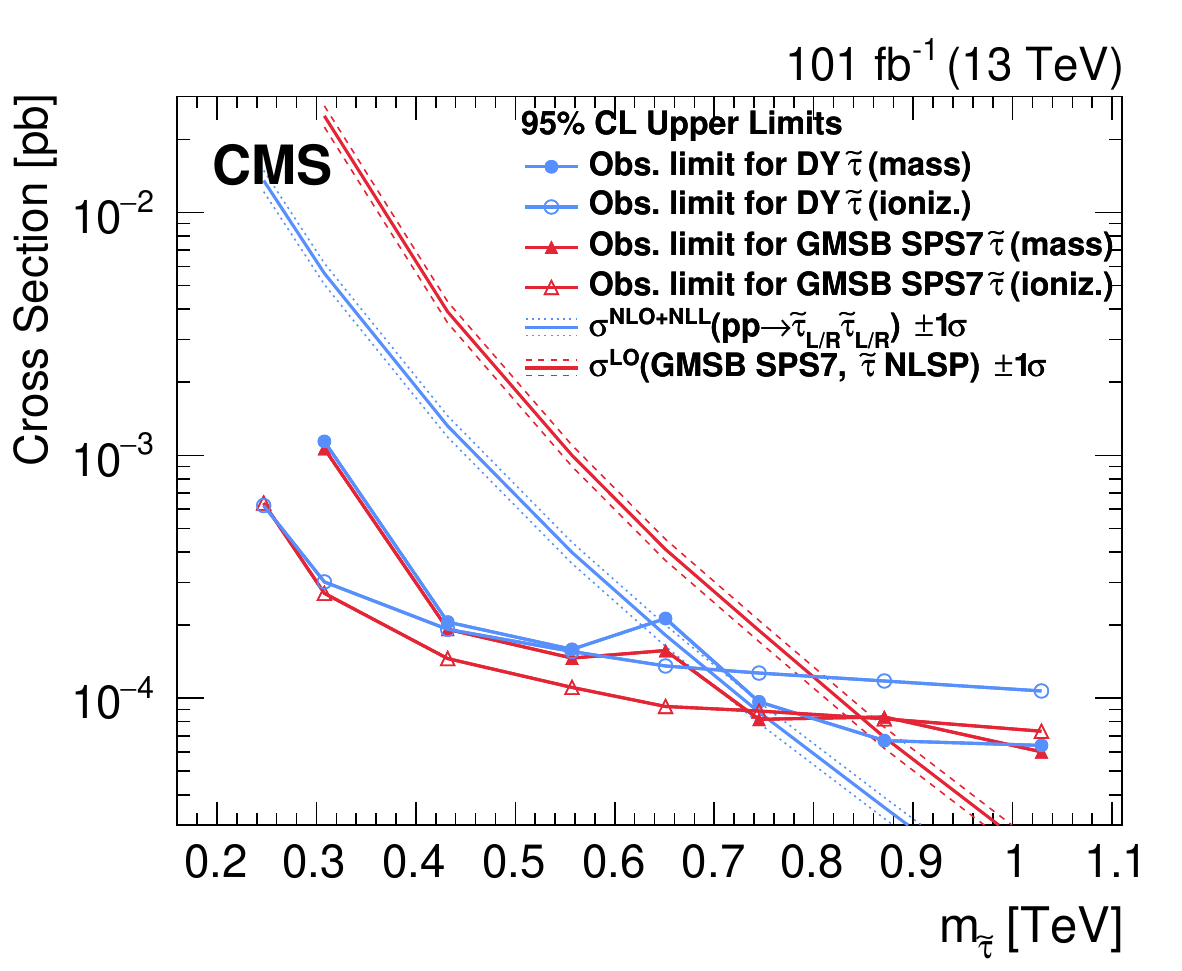}
	\caption{Cross section limits for \PSg (blue circles) and \stopq\ R-hadrons (red triangles) on the left and 
	for the DY pair production of \stau (blue circles) and within the
	{GMSB SPS7} model  (red triangles) on the right. The results obtained with the ionization method are displayed with open symbols, 
	while the symbols for the mass method are filled. Corresponding theoretical predictions are shown using the same color code.}
   \label{fig:massLim1}
\end{figure}

Figure~\ref{fig:massLim1} (left) shows in blue the observed upper limits on the signal production cross section
for the pair production of \PSg \rhadrons, as a function of their mass. 
The cross section limits for the pair production of \stopq\ R-hadrons are displayed in red. 
The theoretical predictions and uncertainties for these simplified models are computed at next-to-next-to-leading order with next-to-next-to-leading logarithm resummation (NNLO+NNLL)~\cite{Beenakker_2016,Borschensky:2014cia,Butterworth:2015oua}. 
At 95\% \CL, this search is able to exclude \PSg R-hadrons with a mass up to 2.06\TeV  
for the ionization method and up to  2.08\TeV  for the mass method.
The ionization (mass) method excludes \stopq\ masses below 1.40 (1.47)\TeV.

This analysis is sensitive to lepton-like HSCP candidates. Limits are set on the two different \stau production models:
direct pair production of \stau and production within the {GMSB SPS7} model. The observed limits are presented in the right panel of Fig.~\ref{fig:massLim1}.
No mass limit can be derived below 300\GeV because of the bias correction applied in the mass method.
The theoretical cross sections for direct pair production depend on the \stau mixing parameters.
However, the signal efficiency is independent of this mixing as we assume a stable \stau. 
Depending on the \stau sector parameters, we are able to set 
three different mass limits for the ionization method (mass method): 0.52 (0.51)\TeV for pair production of \stauR, 
0.64 (0.61)\TeV for pair production of \stauL, and 0.69 (0.64)\TeV if the \stauL and \stauR are mass degenerate.
Figure~\ref{fig:massLim1} (right) shows only
the theoretical prediction for the last case (in blue), including both \stauL\ and \stauR.
In the context of the GMSB SPS7  model, \stau  masses are excluded  up to 0.85\TeV for both the ionization
and mass methods.

\begin{figure}[ht!]
   \centering
	\includegraphics[width=.42\textwidth]{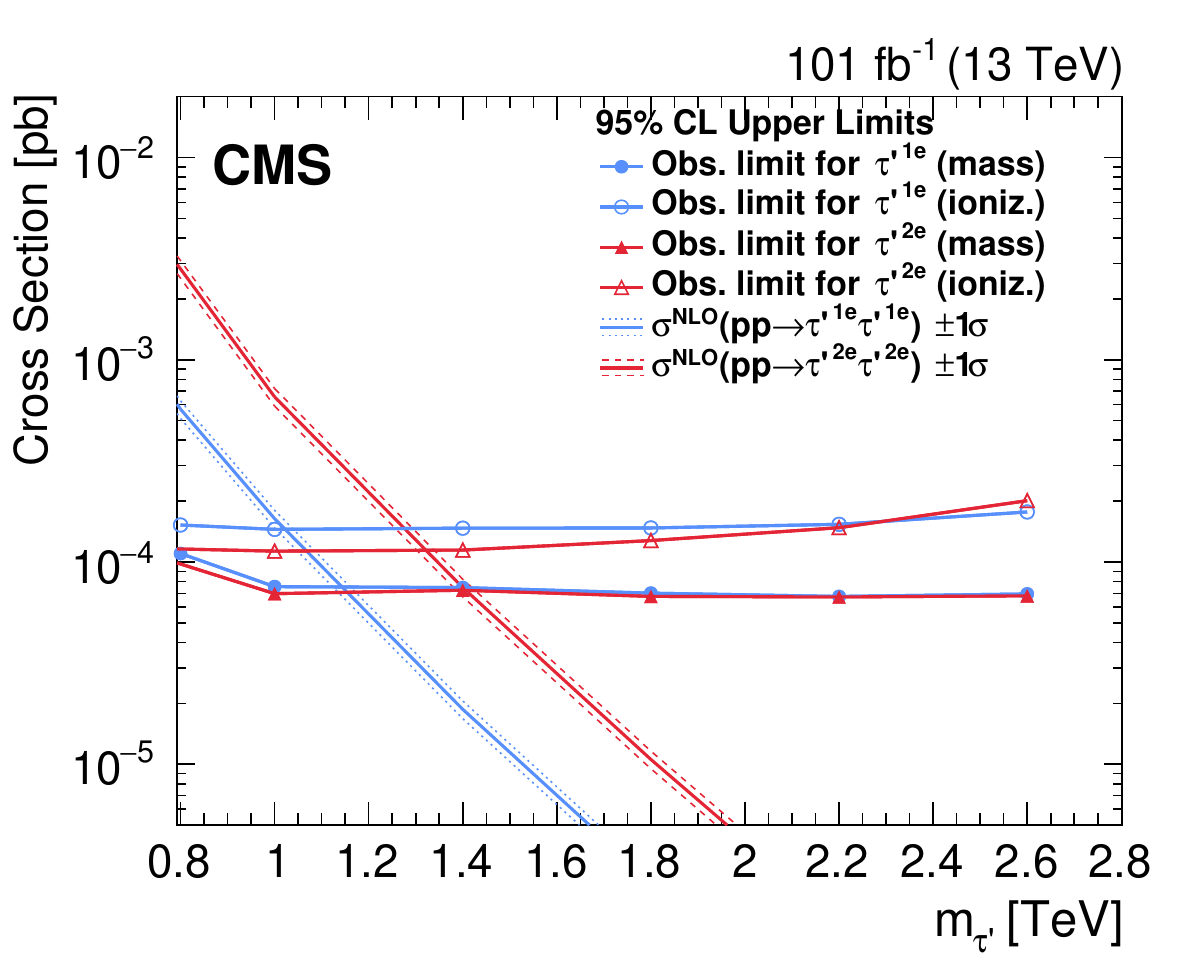}
       \includegraphics[width=.42\textwidth]{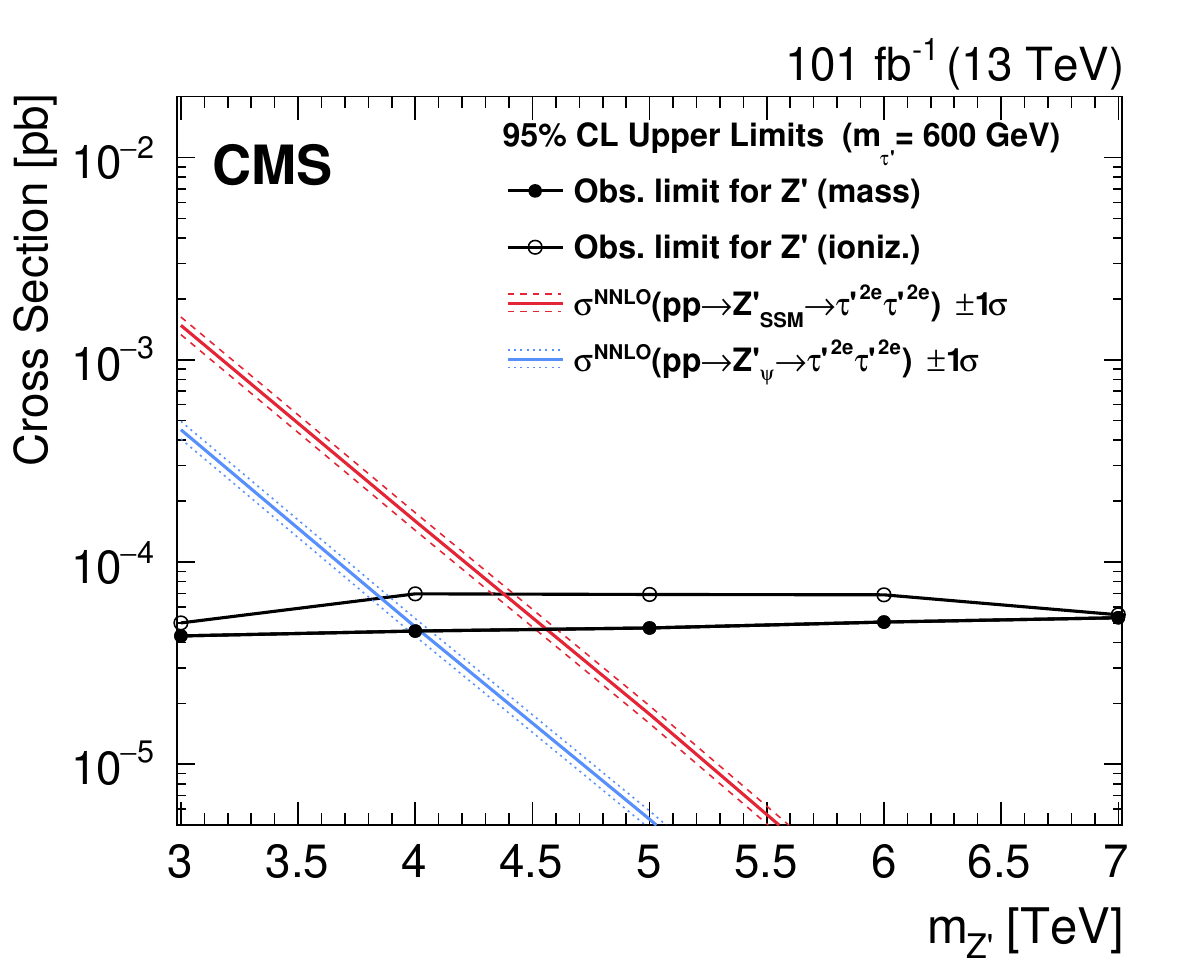}
   \caption{Cross section  limits on the left for the {DY}-produced  \PGtpr  fermions with ${Q} = 1e$ (blue circles) and ${Q} = 2e$ (red triangles), 
	and on the right for the production of \PZpr boson decaying into a pair of \PGtpr fermions of charge $2e$ (black circles).
	The results obtained with the ionization method are displayed with open symbols,
        while the symbols for the mass method are filled.
	The corresponding theoretical predictions for the two {DY}-productions are shown using the same color code as for the limits, depending on the  \PGtpr  charge.
        For the \PZpr production, all the samples assume a narrow width.
        The branching fraction for the \PZpr boson decay to $\PGtpr\PGtpr$ is 1 and a fixed \PGtpr mass of 600\GeV is used.
	The blue (red) curves on the right plot shows the theoretical production cross section for a \ZPrimepsi (\ZPrimeSSM) boson~\cite{Accomando:2010fz} (\cite{HEWETT1989193}).}
	\label{fig:massLim2}
\end{figure}

Figure~\ref{fig:massLim2} (left) presents the limits for the DY production of a pair of  \PGtpr leptons, 
assuming two different electric charge scenarios, ${Q} = 1e$ and $2e$, respectively. Figure~\ref{fig:massLim2} (right) shows the limits for the production cross section of  a \PZpr boson decaying into a pair of \PGtpr fermions of charge $2e$ and a mass of 600\GeV.
The theoretical cross section of production of a \ZPrimepsi (\ZPrimeSSM) boson is taken from~\cite{Accomando:2010fz} (\cite{HEWETT1989193}).
Drell--Yan signals with ${Q} = 1e$ and $2e$ are excluded below 
 1.02 (1.14)\TeV  and 1.32 (1.41)\TeV for the ionization approach (for the mass method), respectively.

The mass limits obtained at  $\sqrt{s}=13$\TeV for various HSCP signal models are summarized in Table~\ref{tab:MassLimits}.  
The observed limits on the mass are close to the expected ones.
The mass method provides slightly better limits than the ionization method
 in the large mass regime,
while the ionization method is more
sensitive in the low mass regime.
These results improve the exclusion mass limits compared to all previous publications.

\begin{table}[ht!]
  \topcaption{Expected and observed mass limits obtained for various HSCP candidate models,
 for the two background estimate methods.}
   \label{tab:MassLimits} 
  \centering
  \begin{tabular}{lcccc} 
     Model       & \multicolumn{2}{c}{Ionization method} & \multicolumn{2}{c}{Mass method} \\
				       & Exp. ({\TeVns}) & Obs. ({\TeVns}) & Exp. ({\TeVns})  & Obs. ({\TeVns}) \\ 
                   \hline
	  {\PSg}                                     & 2.06 $\pm$ 0.06 & 2.06  & 2.08 $\pm$ 0.02  & 2.08  \\
          {\stopq}                                   & 1.43 $\pm$ 0.05 & 1.40  & 1.47 $\pm$ 0.02  & 1.47  \\
	  {GMSB SPS7 \stau}                               & 0.86 $\pm$ 0.07 & 0.85  & 0.87 $\pm$ 0.05  & 0.85  \\
          {pair-prod. \stauR}                        & 0.53 $\pm$ 0.03 & 0.52  & 0.50 $\pm$ 0.07  & 0.51  \\
          {pair-prod. \stauL}                        & 0.66 $\pm$ 0.04 & 0.64  & 0.67 $\pm$ 0.06  & 0.61  \\
          {pair-prod. \stauLR}                       & 0.71 $\pm$ 0.04 & 0.69  & 0.75 $\pm$ 0.08  & 0.64  \\
          {\PGtpr ($Q=1e$) from DY prod.}        & 1.05 $\pm$ 0.05 & 1.02  & 1.14 $\pm$ 0.03  & 1.14  \\
          {\PGtpr ($Q=2e$) from DY prod.}        & 1.35 $\pm$ 0.05 & 1.32  & 1.41 $\pm$ 0.02  & 1.41  \\
          {$\ZPrimepsi \to \PGtpr \PGtpr$ }    & 3.99 $\pm$ 0.21 & 3.95  & 4.03 $\pm$ 0.01 & 4.03  \\
          {$\ZPrimeSSM \to \PGtpr \PGtpr$ }    & 4.53 $\pm$ 0.23 & 4.38  & 4.56 $\pm$ 0.01 & 4.57  \\ 
  \end{tabular}
\end{table}

\begin{figure}[ht!]
    \centering
   \includegraphics[height=.36\textwidth]{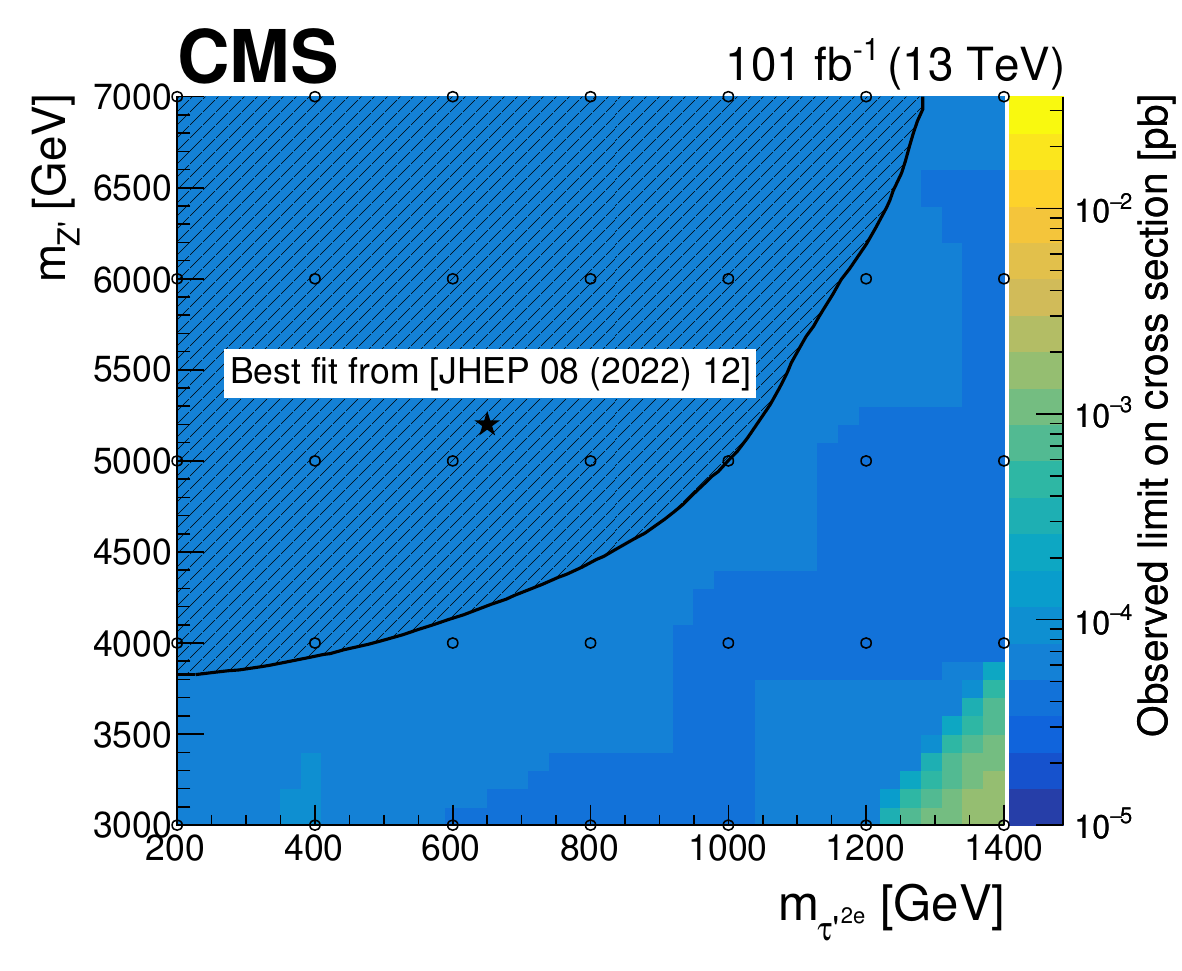}  
    \includegraphics[height=.36\textwidth]{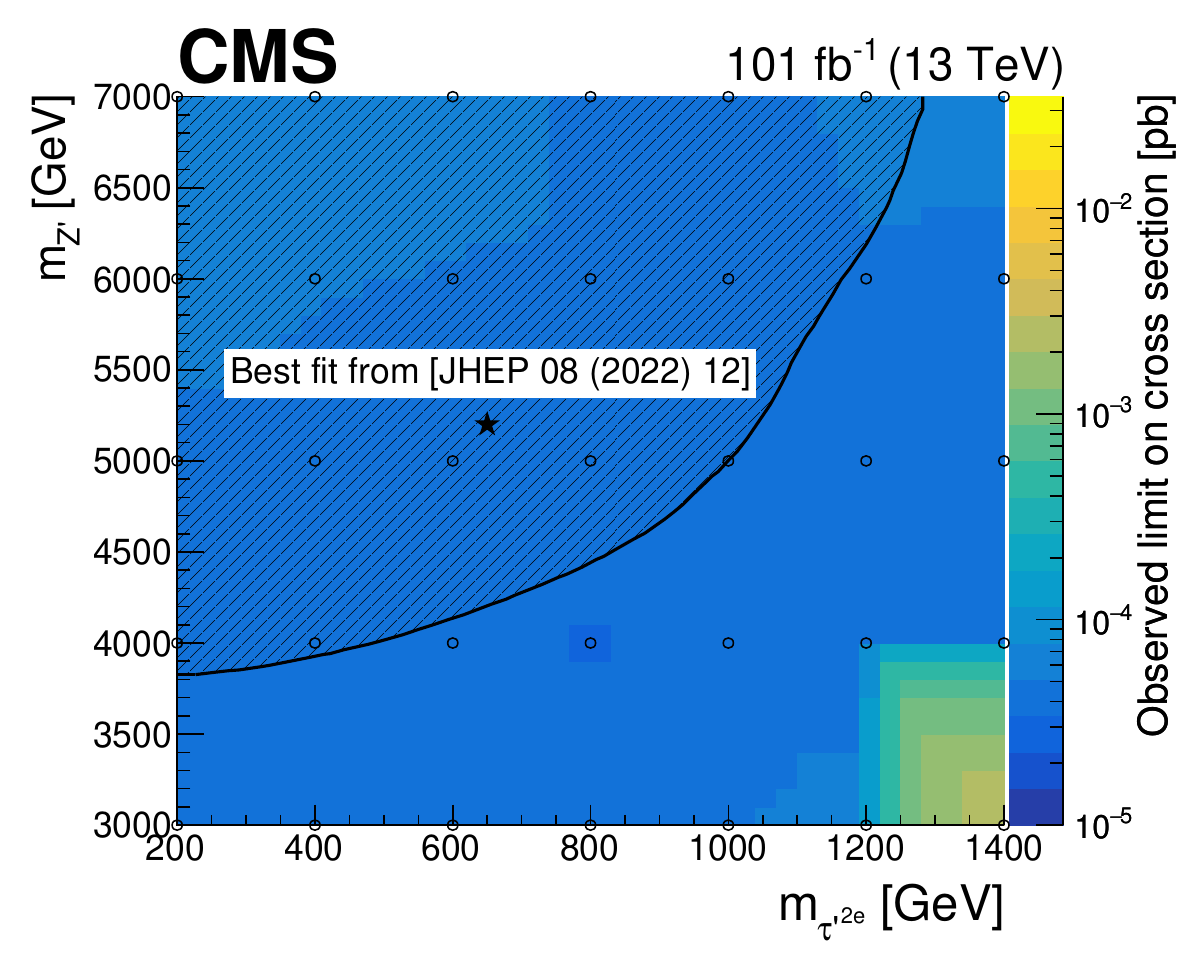}
	 \caption{The two-dimensional exclusion showing the observed cross section limit as a function of the masses of the \PGtpr (on the $x$ axis) and of the \PZpr boson (on the $y$ axis), for the  ionization method on the left and for the mass method on the right. The area above the black solid line corresponds to 
	 the region that is compatible with the {ATLAS} excess from Ref.~\cite{Giudice_2022} and the black star corresponds to the best fit of the ATLAS excess with this model. The empty circles correspond to the 35 simulated mass points.}
    \label{fig:massLimZPrimeTauPrime}
\end{figure}

The results for doubly charged \PGtpr through \PZpr boson resonant production are shown in Fig.~\ref{fig:massLimZPrimeTauPrime} as a 2D 
exclusion plot, taking the signal cross section for the \PZpr from \ZPrimepsi\ models \cite{Accomando:2010fz,atlas_Zprime_2017} and assuming a branching fraction of 
$\PZpr \to \PGtpr \PGtpr$ of 100\%. This plot displays the observed cross section limit as a function of the
\PGtpr mass (on the $x$ axis) and of the \PZpr boson  mass (on the $y$ axis).
The empty circles correspond to the 35 simulated mass points.
The cross section for the full plane is interpolated based on these 35 points. Where the \PZpr boson mass is 3000\GeV and each \PGtpr 
has a mass of 1400\GeV, 
the momentum available to the  \PGtpr is very small, which makes the search less sensitive and results in higher excluded cross sections.
The area above the black solid line in Fig.~\ref{fig:massLimZPrimeTauPrime} corresponds to the region that is compatible with the {ATLAS} excess 
from Ref.~\cite{Giudice_2022}, while the black star corresponds to the best fit of the ATLAS excess with this model.

As the closest mass point to the star is at $m_{\PGtpr}=600\GeV$, 
the observed exclusion limit for that particular $m_{\PGtpr}$ hypothesis is displayed  as a function of the \PZpr boson  mass
in Fig.~\ref{fig:massLim2} (right). Limits can be set on two \PZpr models. 
The exclusion limits for the \ZPrimepsi\ model
lead to an observed mass exclusion at 3.95\TeV for the ionization method and at 4.03\TeV for the mass method.
When the same signal simulation is re-interpreted for a \PZpr boson in the sequential SM (SSM)~\cite{HEWETT1989193},
the observed mass limit is at 4.38\TeV for the ionization method and at 4.57\TeV for the mass method.
The improvement with respect to the \ZPrimepsi\ model is a consequence of the higher theoretical cross section values of the  \ZPrimeSSM\ model. 
The main differences between these two models are the coupling constants to the {SM} particles and the width of the \PZpr boson. 
The width-to-mass ratio of the \PZpr boson changes from 0.53\% in the  \ZPrimepsi\ model to 2.9\% in the  \ZPrimeSSM\ model.
Since these limits are set in the context of a new model, there is no direct comparison to previous results.

As reported in Table~5 of Ref.~\cite{ATLAS_excess}, the maximum excess observed by the ATLAS Collaboration
is obtained for the hypothesis of a pair production of long-lived particles with a mass of 1.4\TeV. This excess corresponds to an
observed signal cross section of $0.0453\unit{fb} / (A \epsilon)$.
Using the acceptance ($A$) and efficiency ($\epsilon$) reported in the associated HEPData record~\cite{ATLAS_excess},
this excess corresponds to an observed signal cross section of 0.59\unit{fb} under the hypothesis of a stable \PSg.  
The observed excluded cross section obtained by the mass method is 0.43\unit{fb} for the same hypothesis.

In Ref.~\cite{Giudice_2022}, an interpretation for the production of a \PZpr boson decaying into a pair of
doubly charged heavy leptons shows that masses of about 5\TeV for the \PZpr boson and 600\GeV for the heavy lepton could explain the ATLAS excess.
For this hypothesis, the present result excludes at the 95\% \CL signal cross sections above 0.047\unit{fb}, and therefore does not confirm the presence of a signal excess
when its corresponding $(A \epsilon)$ is lower than 95\%.

\section{Summary}
\label{sec:summary}
A dedicated search for heavy long-lived charged particles produced in proton-proton collisions at $\sqrt{s} = 13\TeV$ using the CMS detector has been presented.
Since the experimental signature is a highly ionizing isolated track of high \pt, the analysis is based on anomalous \dedx deposits 
in the silicon tracker. 
The data, corresponding to an integrated luminosity of 101\fbinv, are compatible with the background predictions obtained with two different methods. The ionization method is based on the newly introduced ionization variable \fpix\ , which uses only pixel detector information, and the \gstrip\ discriminant, which uses charges collected in the silicon strip detector.  The mass method is based on the
reconstruction of mass using the ionization variable \Ih, providing an estimate of the most probable value of \dedx  of the track, and the momentum  of the candidate.
The two methods lead to similar sensitivity. Cross section limits are set in the context of several models predicting 
the pair production of gluino \rhadrons, top squark \rhadrons, tau sleptons, and \PGtpr leptons with an electric charge 
equal to $1e$ or $2e$.
Gluino \rhadrons\ are excluded at the 95\% \CL with a mass up to 2.08\TeV, top squark \rhadrons\ up to 1.47\TeV, pair produced \stauLR up to 0.69\TeV,
and \PGtpr fermions with an electric charge equal to $1e$ ($2e$) up to 1.14 (1.41)\TeV.
The model predicting a pair of doubly-charged \PGtpr fermions from the \PZpr boson decay  motivated by
an excess reported by the ATLAS Collaboration~\cite{ATLAS_excess} is directly addressed, and no significant deviation from the SM is observed. 
Cross section limits are extracted as a function of the \PZpr and \PGtpr masses.
For a \PGtpr mass of 600\GeV, \ZPrimepsi (\ZPrimeSSM) bosons are excluded at 95\% \CL with a mass up to 4.03 (4.57)\TeV.
The observed limits for stable pair produced tau sleptons, stable gluinos, and stable top squarks, as well as the Drell--Yan \PGtpr and $\PZpr\to\PGtpr\PGtpr$ signals are the best published to date.

\begin{acknowledgments}
   We congratulate our colleagues in the CERN accelerator departments for the excellent performance of the LHC and thank the technical and administrative staffs at CERN and at other CMS institutes for their contributions to the success of the CMS effort. In addition, we gratefully acknowledge the computing centers and personnel of the Worldwide LHC Computing Grid and other centers for delivering so effectively the computing infrastructure essential to our analyses. Finally, we acknowledge the enduring support for the construction and operation of the LHC, the CMS detector, and the supporting computing infrastructure provided by the following funding agencies: SC (Armenia), BMBWF and FWF (Austria); FNRS and FWO (Belgium); CNPq, CAPES, FAPERJ, FAPERGS, and FAPESP (Brazil); MES and BNSF (Bulgaria); CERN; CAS, MoST, and NSFC (China); MINCIENCIAS (Colombia); MSES and CSF (Croatia); RIF (Cyprus); SENESCYT (Ecuador); ERC PRG, RVTT3 and MoER TK202 (Estonia); Academy of Finland, MEC, and HIP (Finland); CEA and CNRS/IN2P3 (France); SRNSF (Georgia); BMBF, DFG, and HGF (Germany); GSRI (Greece); NKFIH (Hungary); DAE and DST (India); IPM (Iran); SFI (Ireland); INFN (Italy); MSIP and NRF (Republic of Korea); MES (Latvia); LMTLT (Lithuania); MOE and UM (Malaysia); BUAP, CINVESTAV, CONACYT, LNS, SEP, and UASLP-FAI (Mexico); MOS (Montenegro); MBIE (New Zealand); PAEC (Pakistan); MES and NSC (Poland); FCT (Portugal); J MESTD (Serbia); MCIN/AEI and PCTI (Spain); MOSTR (Sri Lanka); Swiss Funding Agencies (Switzerland); MST (Taipei); MHESI and NSTDA (Thailand); TUBITAK and TENMAK (Turkey); NASU (Ukraine); STFC (United Kingdom); DOE and NSF (USA).
   
   \hyphenation{Rachada-pisek} Individuals have received support from the Marie-Curie program and the European Research Council and Horizon 2020 Grant, contract Nos.\ 675440, 724704, 752730, 758316, 765710, 824093, 101115353, 101002207, and COST Action CA16108 (European Union); the Leventis Foundation; the Alfred P.\ Sloan Foundation; the Alexander von Humboldt Foundation; the Science Committee, project no. 22rl-037 (Armenia); the Belgian Federal Science Policy Office; the Fonds pour la Formation \`a la Recherche dans l'Industrie et dans l'Agriculture (FRIA-Belgium); the F.R.S.-FNRS and FWO (Belgium) under the ``Excellence of Science -- EOS" -- be.h project n.\ 30820817; the Beijing Municipal Science \& Technology Commission, No. Z191100007219010 and Fundamental Research Funds for the Central Universities (China); the Ministry of Education, Youth and Sports (MEYS) of the Czech Republic; the Shota Rustaveli National Science Foundation, grant FR-22-985 (Georgia); the Deutsche Forschungsgemeinschaft (DFG), among others, under Germany's Excellence Strategy -- EXC 2121 ``Quantum Universe" -- 390833306, and under project number 400140256 - GRK2497; the Hellenic Foundation for Research and Innovation (HFRI), Project Number 2288 (Greece); the Hungarian Academy of Sciences, the New National Excellence Program - \'UNKP, the NKFIH research grants K 131991, K 133046, K 138136, K 143460, K 143477, K 146913, K 146914, K 147048, 2020-2.2.1-ED-2021-00181, and TKP2021-NKTA-64 (Hungary); the Council of Science and Industrial Research, India; ICSC -- National Research Center for High Performance Computing, Big Data and Quantum Computing and FAIR -- Future Artificial Intelligence Research, funded by the NextGenerationEU program (Italy); the Latvian Council of Science; the Ministry of Education and Science, project no. 2022/WK/14, and the National Science Center, contracts Opus 2021/41/B/ST2/01369 and 2021/43/B/ST2/01552 (Poland); the Funda\c{c}\~ao para a Ci\^encia e a Tecnologia, grant CEECIND/01334/2018 (Portugal); the National Priorities Research Program by Qatar National Research Fund;  MCIN/AEI/10.13039/501100011033, ERDF ``a way of making Europe", and the Programa Estatal de Fomento de la Investigaci{\'o}n Cient{\'i}fica y T{\'e}cnica de Excelencia Mar\'{\i}a de Maeztu, grant MDM-2017-0765 and Programa Severo Ochoa del Principado de Asturias (Spain); the Chulalongkorn Academic into Its 2nd Century Project Advancement Project, and the National Science, Research and Innovation Fund via the Program Management Unit for Human Resources \& Institutional Development, Research and Innovation, grant B39G670016 (Thailand); the Kavli Foundation; the Nvidia Corporation; the SuperMicro Corporation; the Welch Foundation, contract C-1845; and the Weston Havens Foundation (USA).
\end{acknowledgments}

\bibliography{auto_generated}
\cleardoublepage \appendix\section{The CMS Collaboration \label{app:collab}}\begin{sloppypar}\hyphenpenalty=5000\widowpenalty=500\clubpenalty=5000\input{EXO-18-002-public-authorlist.tex}\end{sloppypar}
\end{document}

%% file: EXO-18-002-public-authorlist.tex
\cmsinstitute{Yerevan Physics Institute, Yerevan, Armenia}
{\tolerance=6000
A.~Hayrapetyan, A.~Tumasyan\cmsAuthorMark{1}\cmsorcid{0009-0000-0684-6742}
\par}
\cmsinstitute{Institut f\"{u}r Hochenergiephysik, Vienna, Austria}
{\tolerance=6000
W.~Adam\cmsorcid{0000-0001-9099-4341}, J.W.~Andrejkovic, L.~Benato\cmsorcid{0000-0001-5135-7489}, T.~Bergauer\cmsorcid{0000-0002-5786-0293}, S.~Chatterjee\cmsorcid{0000-0003-2660-0349}, K.~Damanakis\cmsorcid{0000-0001-5389-2872}, M.~Dragicevic\cmsorcid{0000-0003-1967-6783}, P.S.~Hussain\cmsorcid{0000-0002-4825-5278}, M.~Jeitler\cmsAuthorMark{2}\cmsorcid{0000-0002-5141-9560}, N.~Krammer\cmsorcid{0000-0002-0548-0985}, A.~Li\cmsorcid{0000-0002-4547-116X}, D.~Liko\cmsorcid{0000-0002-3380-473X}, I.~Mikulec\cmsorcid{0000-0003-0385-2746}, J.~Schieck\cmsAuthorMark{2}\cmsorcid{0000-0002-1058-8093}, R.~Sch\"{o}fbeck\cmsAuthorMark{2}\cmsorcid{0000-0002-2332-8784}, D.~Schwarz\cmsorcid{0000-0002-3821-7331}, M.~Sonawane\cmsorcid{0000-0003-0510-7010}, W.~Waltenberger\cmsorcid{0000-0002-6215-7228}, C.-E.~Wulz\cmsAuthorMark{2}\cmsorcid{0000-0001-9226-5812}
% ...
\par}
\vskip\cmsinstskip
\dag:~Deceased\\
$^{1}$Also at Yerevan State University, Yerevan, Armenia\\
$^{2}$Also at TU Wien, Vienna, Austria\\
% ...
